\documentclass[amsmath,amssymb]{amsart}

\usepackage{graphicx}
\usepackage{revsymb}
\usepackage{dcolumn}
\usepackage{bm}

\begin{document}
\title{Monotonically convergent optimal control theory of quantum systems under a nonlinear interaction with the control field}
\author{M. Lapert$^1$, R. Tehini$^1$, G. Turinici$^2$, D. Sugny$^1$}
\address{$^1$ Institut Carnot de
Bourgogne, UMR 5209 CNRS-Universit\'e de Bourgogne, BP 47870,
21078 Dijon, France}
\address{$^2$ CEREMADE, Universit\'e Paris Dauphine, Place du
Mar\'echal De Lattre De Tassigny, 75775 Paris Cedex 16, France}
\date{\today}
\begin{abstract}
We consider the optimal control of quantum systems interacting
non-linearly with an electromagnetic field. We propose new
monotonically convergent algorithms to solve the optimal
equations. The monotonic behavior of the algorithm is ensured by a
non-standard choice of the cost which is not quadratic in the
field. These algorithms can be constructed for pure and
mixed-state quantum systems. The efficiency of the method is shown
numerically on molecular orientation with a nonlinearity of order
3 in the field. Discretizing the amplitude and the phase of the
Fourier transform of the optimal field, we show that the optimal
solution can be well-approximated by pulses that could be
implemented experimentally.
\end{abstract}
\maketitle
\section{Introduction}
The control of quantum dynamics induced by an intense laser field
continue to be a challenge to both experiment and theory
\cite{warren,rabitz0,nielsen}. In this context, Optimal Control
Theory (OCT) is an efficient tool for designing laser pulses able
to control quantum processes
\cite{zhu0,maday,schirmer,ohtsuki1,ohtsuki2,sugnybook,sugnynew,gross,nakagami,potz1,potz2}.
By construction, the optimal field is the field steering a
dynamical system from the initial state to a desired target state
and minimizing a cost functional which generally penalizes the
energy or the duration of the field. Different methods have been
developed to solve the optimal equations \cite{gross,bryson}.
Among others, monotonically convergent iterative schemes proposed
by Tannor et al. \cite{tannor} and Rabitz et al.
\cite{zhu0,zhu1,zhu} have been applied with success to a variety
of physical and chemical processes
\cite{zhu0,sugny7,bifurcating,sugny2,salomon}. These algorithms
have the particularity to guarantee the increase of the cost
functional at each step of iteration. In this paper, we will
consider the Rabitz formulation of iterative algorithms
\cite{zhu1}. First introduced to treat pure-state quantum systems,
these schemes have been extended and applied to mixed-state
quantum systems, dissipative ones \cite{ohtsuki1,ohtsuki2} and
non-markovian dynamics \cite{saalfrank}. A majority of works has
considered a linear interaction between the quantum system and the
electromagnetic field. This linear interaction corresponds, for
molecular systems, to the first order dipolar approximation
(permanent dipole moment). Due to the intensity of the field or to
the particular structure of the problem, some systems need to go
beyond this approximation \cite{friedrich,stolow1,stolow2,tehini}.
A typical example is given by the control of molecular orientation
and alignment of a linear molecule by non-resonant laser pulses
\cite{seideman,seideman1,friedrich,tehini}. When averaging over
the rapid oscillations of the field, one observes that the
permanent dipole moment plays no role in the control of the
dynamics. In this case, molecular alignment and orientation are
obtained via the polarizability and the hyperpolarizability terms
of the interaction Hamiltonian (see~\cite{oberwolfach} for
information on the controllability of these systems). From a
methodological point of view, the natural question arises of
whether one can apply monotonically convergent algorithms to such
systems interacting non-linearly with the field.

The goal of this work is to answer this question by proposing new
monotonic algorithms when an arbitrary nonlinearity is considered.
A key ingredient to ensure the monotonic convergence of the
algorithms is to consider a non-standard cost functional which
instead of penalizing the intensity of the field, i.e., the square
of the electric field penalizes a higher exponent which depends on
the order of the non-linearity. Note that a similar question has
been treated in \cite{nakagami}. A family of algorithms different
from those proposed in this paper has been developed. In
algorithms of \cite{nakagami}, the cost is quadratic in the field
and the control is decomposed into $n$ components for a
nonlinearity of order $n$. Thus, for each iteration of the
algorithm, $2n$ numerical resolutions of the time-dependent
Schr\"odinger equation are required: $n$ for the wave function and
$n$ for the Lagrange multiplier. On the contrary, in this work, we
use only one component for the control field but at the price of
modifying the cost functional. We construct monotonically
convergent algorithms for pure and mixed-state quantum systems but
they can be generalized straightforwardly to dissipative dynamics.
We test the efficiency of these algorithms on the orientation
dynamics of a linear molecule with non-linearity of order 3
corresponding to the hyperpolarizability terms of the molecule
\cite{tehini}. We use as target states the states which maximize
the orientation in a finite-dimensional restriction of the Hilbert
space. Several works have pointed out the role of these target
states which both optimize the field-free orientation and its
duration \cite{sugny6,sugny3,sugny4,sugny1}. Promising results
have been obtained both for pure and mixed-state quantum systems
corresponding to zero and non-zero temperatures.

Finally, we also analyze the structure of the Fourier transform of
optimal control pulses. Our aim is to show that the optimal
solutions can be well-approximated by pulses that could be
implemented experimentally
\cite{judson,science1,science2,science3}. Such pulses, tailored by
genetic algorithms, have been successfully applied for
experimentally and theoretically controlling different molecular
processes \cite{science1,science2,science3,shir,hertz1,hertz2}. In
the frequency domain, they are characterized by the fact that both
the amplitude and the phase of the Fourier transform (but only for
a finite number of frequencies equally distributed over a given
frequency interval) are optimized \cite{shir,hertz1,hertz2}. This
choice corresponds to the types of pulses that can be implemented
by liquid crystal pulse shapers. As an alternative, we use in this
paper the results of our monotonic optimization algorithms to
construct such pulses. Note that we do not adopt filtering
techniques in the iterative algorithm, which have been proposed by
several works (see \cite{gross} and references therein). The idea
consists generally in applying a filter to the control field at
each iteration in order to satisfy spectral constraints. This
filtering has the drawback that it does not generally yield a
monotonic convergence of the algorithm. Instead, we propose to use
a simpler solution. Starting from the optimal solution obtained by
the monotonic algorithm, we discretize the phase and the amplitude
of its Fourier transform into 640 points or less (640 points
correspond to the number of pixels usually used in pulse-shaping
experiments). From this discretization, we then construct a
piecewise constant Fourier transform and a new time-dependent
electric field by an inverse Fourier transform
\cite{hertz1,hertz2}. We finally compare the optimal result and
the one obtained with the discretized field. We show that the
difference between the two results is negligible when the
structure of the optimal field is sufficiently simple or
equivalently when the number of pixels is sufficiently large.

This paper is organized as follows. We first present the model
system in Sec. \ref{sec2a}. We determine in Sec. \ref{sec2b} the
polynomial equation that must be satisfied by the optimal field.
We then outline in Sec. \ref{sec2c} the principle of monotonically
convergent algorithms for nonlinear interaction both in pure and
mixed-state cases. A special attention is paid to the different
choices and to the flexibility of the method. Generalizing the
proofs of Refs. \cite{zhu1} and \cite{salomon}, we show the
monotonic behavior of the algorithms. Section \ref{sec3} is
devoted to the application of these strategies to molecular
orientation. The results are presented at $T=0~\textrm{K}$ (Sec.
\ref{sec3a}) and $T\neq 0~\textrm{K}$ (Sec. \ref{sec3d}) for the
standard case (i.e. with a linear interaction term) and at
$T=0~\textrm{K}$ for the averaging case (Sec. \ref{sec3b}). We
also propose an algorithm well-suited to the simultaneous
optimization of two laser fields. An example is given by the
non-resonant control of molecular orientation by two-color laser
pulses \cite{tehini}. We finally examine in Sec. \ref{sec3c} the
structure of the Fourier transform of the optimal fields.
\section{Optimal Control Theory}\label{sec2}
The goal of this section is to propose monotonically convergent
algorithms suited to quantum systems interacting nonlinearly with
the control field. To simplify the discussion, we consider the
case of pure-state quantum systems. Following Ref. \cite{ohtsuki2}
and the formalism of super-operator, the proof can be
straightforwardly extended to mixed-state quantum systems (see for
that purpose Sec. \ref{sec3d}). Optimal control theory is
invoked in order to maximize the projection onto a target
state, but it could be equivalently defined for maximizing the
expectation value of a given observable. The proof is a
generalization of the standard proof for linear interaction
\cite{zhu1} and of the proof given in Ref. \cite{salomon}.
\subsection{The model system}\label{sec2a}
We consider a quantum system interacting with an electromagnetic
field whose dynamics is governed by the following time-dependent
Schr\"odinger equation
\begin{equation}\label{eq1}
i\frac{\partial}{\partial
t}|\psi(t)\rangle=\hat{H}(t)|\psi(t)\rangle,
\end{equation}
which is written in units such that $\hbar=1$. The Hamiltonian
$\hat{H}(t)$ of the system is given by
\begin{equation}\label{eq2}
\hat{H}(t)=\hat{H}_0-\hat{\mu}E(t)-\hat{\alpha}E(t)^2-\hat{\beta}E(t)^3\cdots,
\end{equation}
where $\hat{H}_0$ is the field-free Hamiltonian. The other terms
describe the interaction between the system and the laser field
$E(t)$. This interaction is written as a polynomial expansion in
$E(t)$ whose coefficients are the operators $\hat{\mu}$,
$\hat{\alpha}$, $\hat{\beta}\cdots$. For a linear molecule
interacting with a linearly polarized laser field, the different
operators $\hat{\mu}$, $\hat{\alpha}$ and $\hat{\beta}$ are
associated to the permanent dipole moment $\mu_0$, the
polarizability components $\alpha_\parallel$ and $\alpha_\perp$
and the hyperpolarizability components $\beta_\parallel$ and
$\beta_\perp$ of the molecule \cite{tehini}. These different
molecular constants will be used in numerical computations of Sec.
\ref{sec3}.
\subsection{Critical point}\label{sec2b}
Let $|\phi_0\rangle$ and $|\phi_f\rangle$ be the initial and the
target states of the control. We denote by $t_f$ the duration of
the control. We define the optimal control theory through the
following cost functional:
\begin{equation}\label{eq3}
J=|\langle \phi_f|\psi(t_f)\rangle|^2-\int_0^{t_f}\lambda
E(t)^{2n}dt,
\end{equation}
where $n$ is a positive integer. The even exponent of the
integrand and the choice $\lambda\geq 0$ ensure the negativity of
the second term of Eq. (\ref{eq3}). $n$ is taken equal to 1 for a
linear interaction but we will see that, in order to obtain
monotonic algorithms, larger values of $n$ have to be considered
when the system interacts non-linearly with the field. $\lambda$
is a penalty factor which weights the importance of the laser
fluence. Following \cite{sundermann}, we will replace in Sec.
\ref{sec3} this constant by $\lambda/s(t)$, where $s(t)=\sin^2(\pi
t/t_f)$, which penalizes more strongly the amplitude of the pulse
at the beginning and at the end of the control. This allows to
obtain more realistic optimal solutions.

We introduce the augmented cost functional $\bar{J}$ which is
defined through the adjoint state $|\chi(t)\rangle$ as follows
\begin{eqnarray}\label{eq4}
\bar{J}=|\langle \phi_f|\psi(t_f)\rangle|^2-\int_0^{t_f}\lambda
E(t)^{2n}dt-2\Im[\langle\psi(t_f)|\phi_f\rangle\int_0^{t_f}\langle\chi(t)|(i\frac{\partial}{\partial
t}-\hat{H})|\psi(t)\rangle dt],
\end{eqnarray}
where $\Im$ denotes the imaginary part. The optimal electric field
is solution of the equation
\begin{equation}\label{eq5}
\frac{\delta \bar{J}}{\delta E(t)}=0
\end{equation}
which is a polynomial equation in $E(t)$:
\begin{eqnarray}\label{eq6}
2n\lambda
E(t)^{2n-1}+2\Im[\langle\psi(t_f)|\phi_f\rangle\langle\chi(t)|\hat{\mu}+2\hat{\alpha}E(t)+3\hat{\beta}E(t)^2|\psi(t)\rangle]=0.
\end{eqnarray}
The second term of Eq. (\ref{eq6}) can be modified by using the
fact that
\begin{equation}\label{eq7}
\frac{d}{dt}\langle\psi(t)|\chi(t)\rangle=0.
\end{equation}
The equation for the optimal field finally reads
\begin{eqnarray}\label{eq8}
2n\lambda
E(t)^{2n-1}+2\Im[\langle\psi(t)|\chi(t)\rangle\langle\chi(t)|\hat{\mu}+2\hat{\alpha}E(t)+3\hat{\beta}E(t)^2|\psi(t)\rangle]=0.
\end{eqnarray}
Setting the variations of $\bar{J}$ with respect to
$|\psi(t)\rangle$ and $|\chi(t)\rangle$ to 0 ensures that
$|\psi(t)\rangle$ and $|\chi(t)\rangle$ satisfy the Schr\"odinger
equation (\ref{eq1}). To summarize, an extremum of $\bar{J}$
satisfies the equations
\begin{eqnarray}\label{eq9}
\begin{array}{ll}
(i\frac{\partial}{\partial t}-\hat{H}(t))|\psi(t)\rangle=0 \\
|\psi(0)\rangle=|\phi_0\rangle
\end{array}
\end{eqnarray}
for the state $|\psi(t)\rangle$ and
\begin{eqnarray}\label{eq10}
\begin{array}{ll}
(i\frac{\partial}{\partial t}-\hat{H}(t))|\chi(t)\rangle=0 \\
|\chi(t_f)\rangle=|\phi_f\rangle
\end{array}
\end{eqnarray}
for the adjoint state $|\chi(t)\rangle$, the control field $E(t)$
being solution of Eq. (\ref{eq8}).

\subsection{Monotonically convergent algorithm}\label{sec2c}
We describe different iterative algorithms to solve the optimal
equations of Sec. \ref{sec2b}. To simplify the presentation of
computations, we consider nonlinearity of order 3 and a cost which
is quartic in the field ($n=2$).

At step $k\geq 1$ of the algorithm, the system is described by the
quadruplet
$(|\psi_k(t)\rangle,|\chi_{k-1}(t)\rangle,E_k(t),\tilde{E}_{k-1}(t))$
where $|\psi_k(t)\rangle$ is the state of the system,
$|\chi_{k-1}(t)\rangle$ the adjoint state, $E_k(t)$ and
$\tilde{E}_{k-1}(t)$ the electric fields associated respectively
to the forward propagation of $|\psi_k(t)\rangle$ and to the
backward propagation of $|\chi_{k-1}(t)\rangle$.
$|\psi_k(t)\rangle$ and $|\chi_{k-1}(t)\rangle$ are solutions of
the following time-dependent Schr\"odinger equations
\begin{equation}\label{eq11}
i\frac{\partial}{\partial
t}|\psi_k(t)\rangle=\hat{H}(E_k)|\psi_k(t)\rangle
\end{equation}
and
\begin{equation}\label{eq12}
i\frac{\partial}{\partial
t}|\chi_{k-1}(t)\rangle=\hat{H}(\tilde{E}_{k-1})|\chi_{k-1}(t)\rangle,
\end{equation}
where
$\hat{H}(E(t))=\hat{H}_0-\hat{\mu}E(t)-\hat{\alpha}E(t)^2-\hat{\beta}E(t)^3$.
For $|\psi_k(t)\rangle$, we impose the initial condition
$|\psi_k(0)\rangle=|\phi_0\rangle$ and for $|\chi_{k-1}(t)\rangle$
the final condition $|\chi_{k-1}(t_f)\rangle=|\phi_f\rangle$. The
iteration is initiated by a trial electric field $E_0(t)$. At step
0 of the algorithm, we propagate forward the state
$|\psi_0(t)\rangle$ with the electric field $E_0(t)$. The cost
functional $J_k$ at step $k$ is defined by
\begin{equation}\label{eq13}
J_k =|\langle\phi_f |\psi_k (t_f)\rangle|^2-\int_0^{t_f} \lambda
E_k^4.
\end{equation}

The algorithm determines the quadruplet
$(|\psi_{k+1}(t)\rangle,|\chi_{k}(t)\rangle,E_{k+1}(t),\tilde{E}_{k}(t))$
at step $k+1$ from the one at step $k$. This is done by requiring
that the variation $\Delta J=J_{k+1}-J_k$ of the cost $J$ from
step $k$ to step $k+1$ is positive and that the limits (if they
exist) of the sequences $(E_k)_{k\in\mathbb{N}}$ and
$(\tilde{E}_k)_{k\in\mathbb{N}}$ are solutions of Eq. (\ref{eq8}).

For that purpose, we introduce the functions $P_{k+1}(t)=|\langle
\chi_k(t)|\psi_{k+1}(t)\rangle|^2$ and
$\tilde{P}_{k+1}(t)=|\langle
\chi_{k+1}(t)|\psi_{k+1}(t)\rangle|^2$. Differentiating with
respect to time these two functions leads to
\begin{eqnarray}\label{eq14}
\frac{d}{dt}P_{k+1}(t)=\mu_{k,k+1}(\tilde{E}_k-E_{k+1})+\alpha_{k,k+1}(\tilde{E}_k^2-E_{k+1}^2)
+\beta_{k,k+1}(\tilde{E}_k^3-E_{k+1}^3)
\end{eqnarray}
for $P_{k+1}$ and to
\begin{eqnarray}\label{eq15}
&
&\frac{d}{dt}\tilde{P}_{k+1}(t)=\mu_{k+1,k+1}(\tilde{E}_{k+1}-E_{k+1})+\alpha_{k+1,k+1}(\tilde{E}_{k+1}^2-E_{k+1}^2)\nonumber
\\& & +\beta_{k+1,k+1}(\tilde{E}_{k+1}^3-E_{k+1}^3)
\end{eqnarray}
for $\tilde{P}_{k+1}$. In Eqs. (\ref{eq14}) and (\ref{eq15}), we
have introduced the notation
\begin{equation}\label{eq16}
A_{k,k'}=2\Im[\langle\psi_{k'}(t)|\chi_{k}(t)\rangle\langle\chi_k(t)|\hat{A}|\psi_{k'}(t)\rangle]
\end{equation}
for a given observable $\hat{A}$. The functions $P_k$ and
$\tilde{P}_k$ fulfill by definition the following relations:
\begin{eqnarray}\label{eq17}
& &\tilde{P}_{k+1}(t_f)=|\langle\phi_f|\psi_{k+1}(t_f)\rangle|^2=P_{k+1}(t_f)\nonumber\\
&
&\tilde{P}_{k+1}(0)=|\langle\chi_{k+1}(0)|\phi_0\rangle|^2=P_{k+2}(0),
\end{eqnarray}
and a direct integration gives
\begin{equation}\label{eq18}
P_k(t_f)=P_k(0)+\int_0^{t_f}\frac{dP_k(t)}{dt}dt.
\end{equation}

The variation $\Delta J$ is given by
\begin{eqnarray}\label{eq19}
\Delta
J=J_{k+1}-J_k=|\langle\phi_f|\psi_{k+1}(t_f)\rangle|^2-|\langle\phi_f|\psi_{k}(t_f)\rangle|^2-\int_0^{t_f}\lambda
[E_{k+1}(t)^4-E_k(t)^4]dt.
\end{eqnarray}
Using the fact that
$|\langle\phi_f|\psi_{k+1}(t_f)\rangle|^2-|\langle\phi_f|\psi_{k}(t_f)\rangle|^2=P_{k+1}(t_f)-P_k(t_f)$
and Eqs. (\ref{eq17}), one deduces that $\Delta J=P_1+P_2$ where
\begin{eqnarray}\label{eq20}
& &P_1=-\int_0^{t_f}\lambda [E_{k+1}^4-\tilde{E}_k^4]+\nonumber\\
&
&\int_0^{t_f}[(\tilde{E}_k-E_{k+1})\mu_{k,k+1}+(\tilde{E}_k^2-E_{k+1}^2)\alpha_{k,k+1}+(\tilde{E}_k^3-E_{k+1}^3)\beta_{k,k+1}]dt
\end{eqnarray}
and
\begin{eqnarray}\label{eq21}
& &P_2=\int_0^{t_f}\lambda [E_{k}^4-\tilde{E}_k^4]-\nonumber\\
&
&\int_0^{t_f}[(\tilde{E}_k-E_{k})\mu_{k,k}+(\tilde{E}_k^2-E_{k}^2)\alpha_{k,k}+(\tilde{E}_k^3-E_{k}^3)\beta_{k,k}]dt.
\end{eqnarray}
To ensure the monotonic behavior of the algorithm, we choose the
fields $E_{k+1}$ and $\tilde{E}_k$ such that the integrals $P_1$
and $P_2$ are positive. A sufficient condition is to impose that
the two integrands $\mathcal{P}_1$ and $\mathcal{P}_2$ associated
to $P_1$ and $P_2$ are positive \cite{salomon}. To be more
precise, we first determine $\tilde{E}_k$ from $E_k$ such that
$P_2$ is positive and then we determine $E_{k+1}$ from
$\tilde{E}_k$ such that $P_1$ is positive. $|\psi_{k+1}\rangle$
and $|\chi_k\rangle$ are computed from a forward and a backward
propagation with the fields $E_{k+1}$ and $\tilde{E}_k$.\\

Starting from these conditions, we introduce two algorithms.\\
\textbf{Algorithm I:}\\
$\mathcal{P}_1$ and $\mathcal{P}_2$ are respectively viewed as
functions of $E_{k+1}$ and $\tilde{E}_k$. $E_{k+1}$ and
$\tilde{E}_k$ are defined as the control fields which maximize
$\mathcal{P}_1$ and $\mathcal{P}_2$. The maxima of these
polynomials are positive since $\mathcal{P}_1(\tilde{E}_k)=0$ and
$\mathcal{P}_2(E_k)=0$. As already mentioned, we first determine
for each time $t$ the maximum of $\mathcal{P}_2$ and then the one
of $\mathcal{P}_1$. The integer $n$ of the cost is chosen
sufficiently large to ensure that the fields which maximize
$\mathcal{P}_1$ and $\mathcal{P}_2$ are finite. This means that we
choose $n$ such that the terms $-\lambda E_{k+1}^{2n}$ and
$-\lambda \tilde{E}_k^{2n}$ are the monomials of higher degree in
$\mathcal{P}_1$ and $\mathcal{P}_2$. We then have:
\begin{eqnarray}\label{eq22}
& &\lim_{E_{k+1}\to \pm \infty} \mathcal{P}_1(E_{k+1})=0\nonumber \\
& &\lim_{\tilde{E}_k\to \pm \infty}
\mathcal{P}_2(\tilde{E}_k)=0\nonumber
\end{eqnarray}
which satisfies the requirement. For nonlinearity of order 3, the
choice $n=2$ is sufficient. Finally, if we assume that the
algorithm converges then we can check that the limits of the
sequences $(E_k)_{k\in\mathbb{N}}$ and
$(\tilde{E}_k)_{k\in\mathbb{N}}$ are solutions of Eq. (\ref{eq8}).
For that, we respectively differentiate $\mathcal{P}_1$ and
$\mathcal{P}_2$ with respect to $E_{k+1}$ and $\tilde{E}_k$. We
next replace in the derivatives of $\mathcal{P}_1$ and
$\mathcal{P}_2$, $E_{k+1}$, $E_k$ and $\tilde{E}_k$ by $E$,
$|\psi_{k+1}\rangle$, $|\psi_k\rangle$ by $|\psi\rangle$ and
$|\chi_k\rangle$ by $|\chi\rangle$. It is then straightforward to
see that the limit $E(t)$ satisfies the optimal equation (\ref{eq8}).\\
\textbf{Algorithm II:}\\
We first write $\mathcal{P}_1$ and $\mathcal{P}_2$ as follows:
\begin{eqnarray}\label{eq23}
&
&\mathcal{P}_1=(E_{k+1}-\tilde{E}_k)[-\lambda(E_{k+1}^3+E_{k+1}^2\tilde{E}_k+
E_{k+1}\tilde{E}_k^2+\tilde{E}_k^3)\\
& &
-\mu_{k,k+1}-\alpha_{k,k+1}(E_{k+1}+\tilde{E}_k)-\beta_{k,k+1}(E_{k+1}^2+\tilde{E}_kE_{k+1}+\tilde{E}_k^2)]\nonumber
,
\end{eqnarray}
and
\begin{eqnarray}\label{eq24}
&
&\mathcal{P}_2=(\tilde{E}_k-E_k)[-\lambda(\tilde{E}_k^3+\tilde{E}_k^2E_k+
E_k^2\tilde{E}_k+E_k^3)\\
& &
-\mu_{k,k}-\alpha_{k,k}(\tilde{E}_k-E_k)-\beta_{k,k}(\tilde{E}_{k}^2+E_k\tilde{E}_{k}+E_k^2)]\nonumber
.
\end{eqnarray}
We then introduce two positive constants $\eta_1$ and $\eta_2$ by
setting:
\begin{eqnarray}\label{eq25}
&
&E_{k+1}-\tilde{E}_k=\eta_1[-\lambda(E_{k+1}^3+E_{k+1}^2\tilde{E}_k+
E_{k+1}\tilde{E}_k^2+\tilde{E}_k^3)\nonumber \\
& &
-\mu_{k,k+1}-\alpha_{k,k+1}(E_{k+1}+\tilde{E}_k)-\beta_{k,k+1}(E_{k+1}^2+\tilde{E}_kE_{k+1}+\tilde{E}_k^2)]
,
\end{eqnarray}
and
\begin{eqnarray}\label{eq26}
& &
\tilde{E}_k-E_k=\eta_2[-\lambda(\tilde{E}_k^3+\tilde{E}_k^2E_k+
E_k^2\tilde{E}_k+E_k^3)\nonumber\\
& &
-\mu_{k,k}-\alpha_{k,k}(\tilde{E}_k+E_k)-\beta_{k,k}(\tilde{E}_{k}^2+E_k\tilde{E}_{k}+E_k^2)]
.
\end{eqnarray}
Equations. (\ref{eq25}) and (\ref{eq26}) are viewed respectively
as equations in $E_{k+1}$ and $\tilde{E}_k$. $E_{k+1}$ and
$\tilde{E}_k$ are defined as one of the solutions of these two
equations. By definition of the constants $\eta_1$ and $\eta_2$,
the values of $\mathcal{P}_1$ and $\mathcal{P}_2$ for these fields
are positive. The integer $n$ is chosen sufficiently large to
ensure that Eqs. (\ref{eq25}) and (\ref{eq26}) always have a real
solution respectively in $E_{k+1}$ and $\tilde{E}_k$. For
nonlinearity of order 3, it is sufficient to take $n=2$. When Eqs.
(\ref{eq25}) and (\ref{eq26}) have more than one real solution at
time $t$, we numerically choose the solution that is  closest to
the one at time $t-dt$ (for the forward propagation) or $t+dt$
(for the backward propagation). The processus is initiated by
imposing that $E_k(0)=0$ and $\tilde{E}_k(t_f)=0$. This allows one
to obtain smooth optimal fields without discontinuity. As for the
algorithm I, we can check that the limits of the sequences
$(E_k)_{k\in\mathbb{N}}$ and $(\tilde{E}_k)_{k\in\mathbb{N}}$
satisfy Eq. (\ref{eq8}). This can be done by replacing
$\tilde{E}_k$ and $E_{k+1}$ by $E$ in Eqs. (\ref{eq25}) and
(\ref{eq26}).

In the two cases, the structure of the algorithms can be
summarized as follows. At step $k+1$, we propagate backward in
time the adjoint state $|\chi_k\rangle$ with the field
$\tilde{E}_k$ determined from $P_2$. We then compute the forward
evolution of $|\psi_{k+1}\rangle$ from $|\phi_0\rangle$. For this
second propagation, we use the field $E_{k+1}$ defined from $P_1$.
Note that a simpler solution which gives a slower convergence
consists in choosing $\tilde{E}_k=E_k$, i.e., to propagate
$|\psi_k\rangle$ and $|\chi_k\rangle$ with the same field.
\section{Control of molecular orientation}\label{sec3}
\subsection{Introduction}
In this section, we investigate the control of orientation
dynamics of a diatomic molecule driven by an electromagnetic field
\cite{seideman,seideman1}. This control is taken as a prototype to
test the efficiency of the algorithm. The application of OCT to
molecular alignment and orientation is relatively recent
\cite{salomon,nakagami,seiddiss3}. One of the main results of Ref.
\cite{salomon} is that the optimal oriented state (see below for a
definition) is reached by rotational ladder climbing, i.e., by
successive rotational excitations. The corresponding optimal pulse
is however very long, of the order of 20 rotational periods, which
could be problematic for practical applications. We consider
shorter durations in this paper of the order of the rotational
period $T_{per}$. We have chosen $t_f=T_{per}$ but other durations
can be considered. Note that for controls much shorter than
$T_{per}$, the optimal solution is very close to the kick
mechanism largely explored using the sudden-impact model
\cite{sugnykick}. The $CO$ molecule is taken as an example. The
units used are atomic units unless otherwise specified.

The molecule is described in a rigid-rotor approximation
interacting with a linearly polarized laser pulse nonresonant with
vibronic frequencies. In this case, the Hamiltonian $\hat{H}$ can
be written as follows \cite{kanai,tehini}
\begin{eqnarray}\label{eq27}
&
&\hat{H}=B\hat{J}^2-\mu_0E(t)\cos\theta-\frac{1}{2}[(\alpha_\parallel-\alpha_\perp)\cos^2\theta+\alpha_\perp]E(t)^2\nonumber
\\& &
-\frac{1}{6}[(\beta_\parallel-3\beta_\perp)\cos^3\theta+3\beta_\perp\cos\theta]E(t)^3,
\end{eqnarray}
where $B$ and $\mu_0$ are the rotational constant and the
permanent dipole moment. $\alpha_\parallel$, $\alpha_\perp$,
$\beta_\parallel$ and $\beta_\perp$ are respectively the
polarizability and the hyperpolarizability components of the
molecule. The labels $\parallel$ and $\perp$ indicate the
components parallel and perpendicular to the internuclear axis.
For the $CO$ molecule, we have chosen the following numerical
values $B=1.9313~\textrm{cm}^{-1}$ and $\mu_0=0.044$,
$\alpha_\parallel=15.65$, $\alpha_\perp=11.73$,
$\beta_\parallel=28.35$ and $\beta_\perp=6.64$ in atomic units
\cite{sekino,maroulis}. $J^2$ is the angular momentum operator and
$\theta$ the angle between the direction of the molecular axis and
the polarization vector. A basis of the Hilbert space is given by
the spherical harmonics $|j,m\rangle$ with $j\geq 0$ and $-j\leq
m\leq j$.
\subsection{Zero rotational temperature}\label{sec3a}
In this section, we consider the limit of zero rotational
temperature. We recall that the expectation value $\langle
\cos\theta\rangle$ is usually taken as a quantitative measure of
orientation \cite{seideman,seideman1}. Here, we replace this
measure by the projection onto a target state $|\phi_f\rangle$. We
consider target states recently introduced for the orientation
which both maximize the field-free orientation and its duration
\cite{sugny3,sugny6}. To construct this target state, we restrict
the Hilbert space to a finite-dimensional one defined by a maximum
value of $j$ denoted $j_{opt}$. For $CO$, we have chosen
$j_{opt}=4$ which leads to a maximum of $\langle\cos\theta\rangle$
of the order of 0.9. In this reduced Hilbert space, the operator
$\cos\theta$ has a non-degenerate discrete spectrum. The target
state $|\phi_f\rangle$ is then defined as the eigenvector of
$\cos\theta$ of highest eigenvalue. The initial state is the state
$|0,0\rangle$. We also recall that the projection $m$ of the
angular momentum $j$ on the field polarization axis is a conserved
quantum number due to cylindrical symmetry.

We now apply the monotonically convergent algorithms I and II.
\begin{figure}
\includegraphics[height=1.75in]{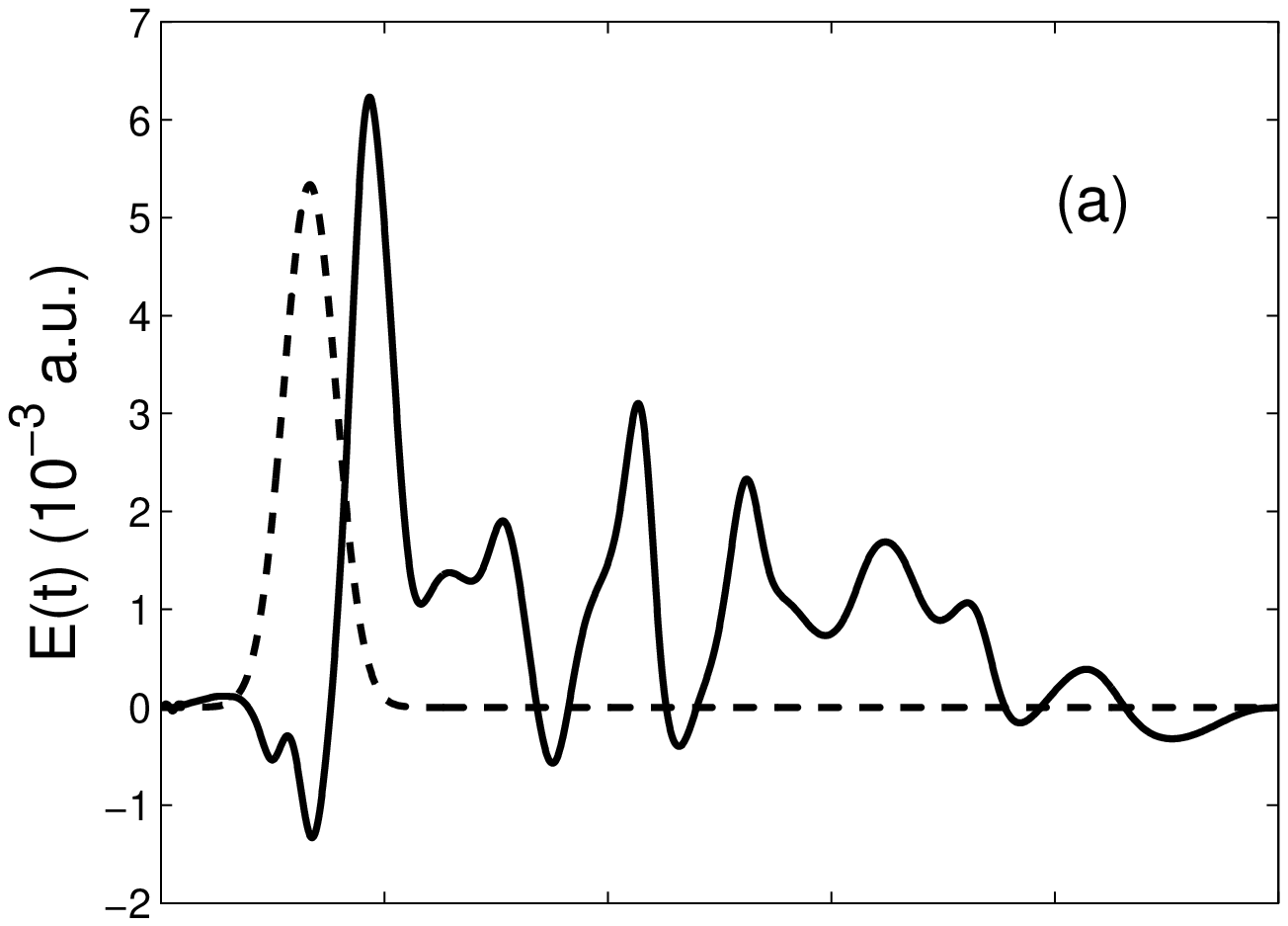}
\includegraphics[height=1.75in]{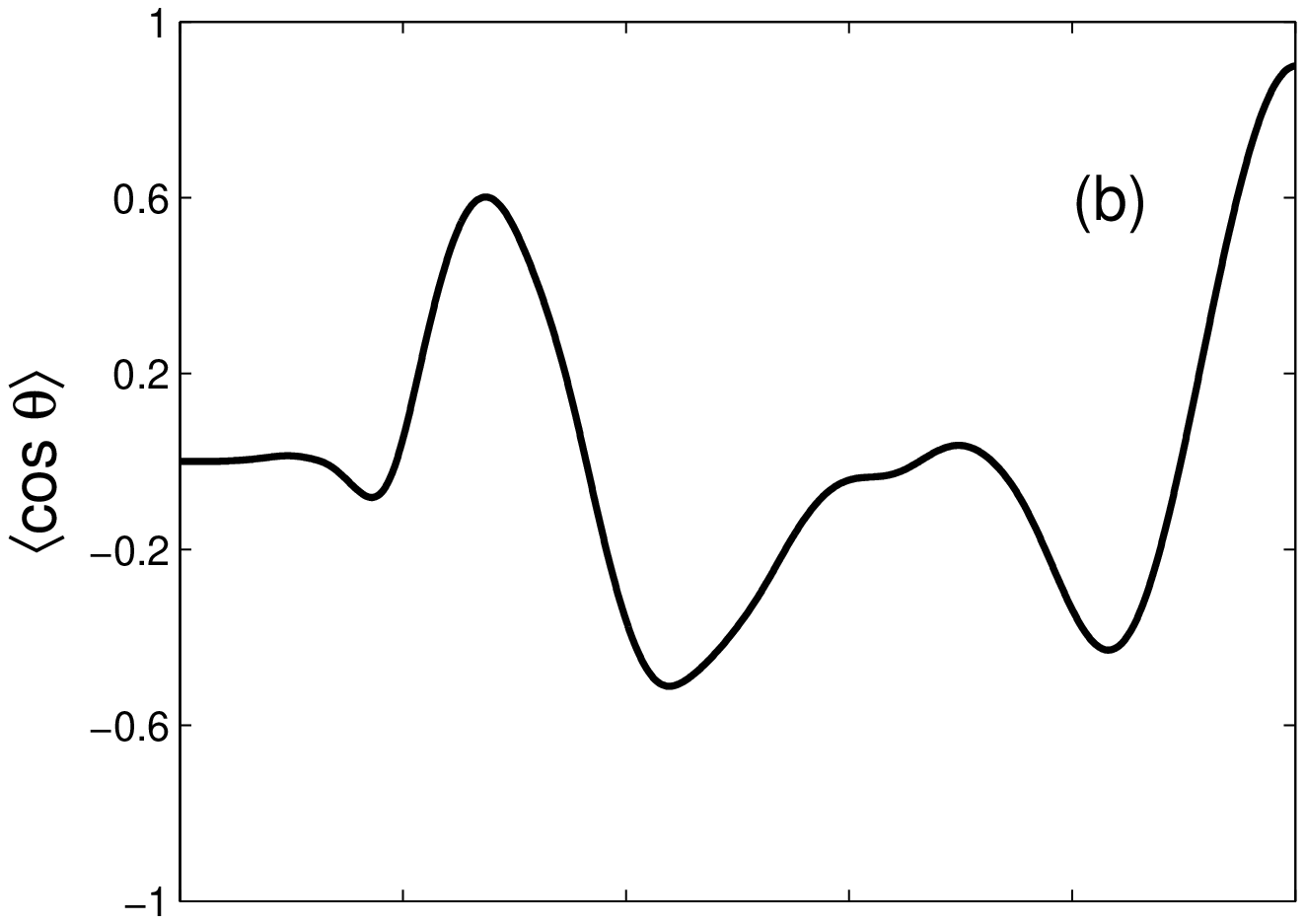}
\includegraphics[height=1.75in]{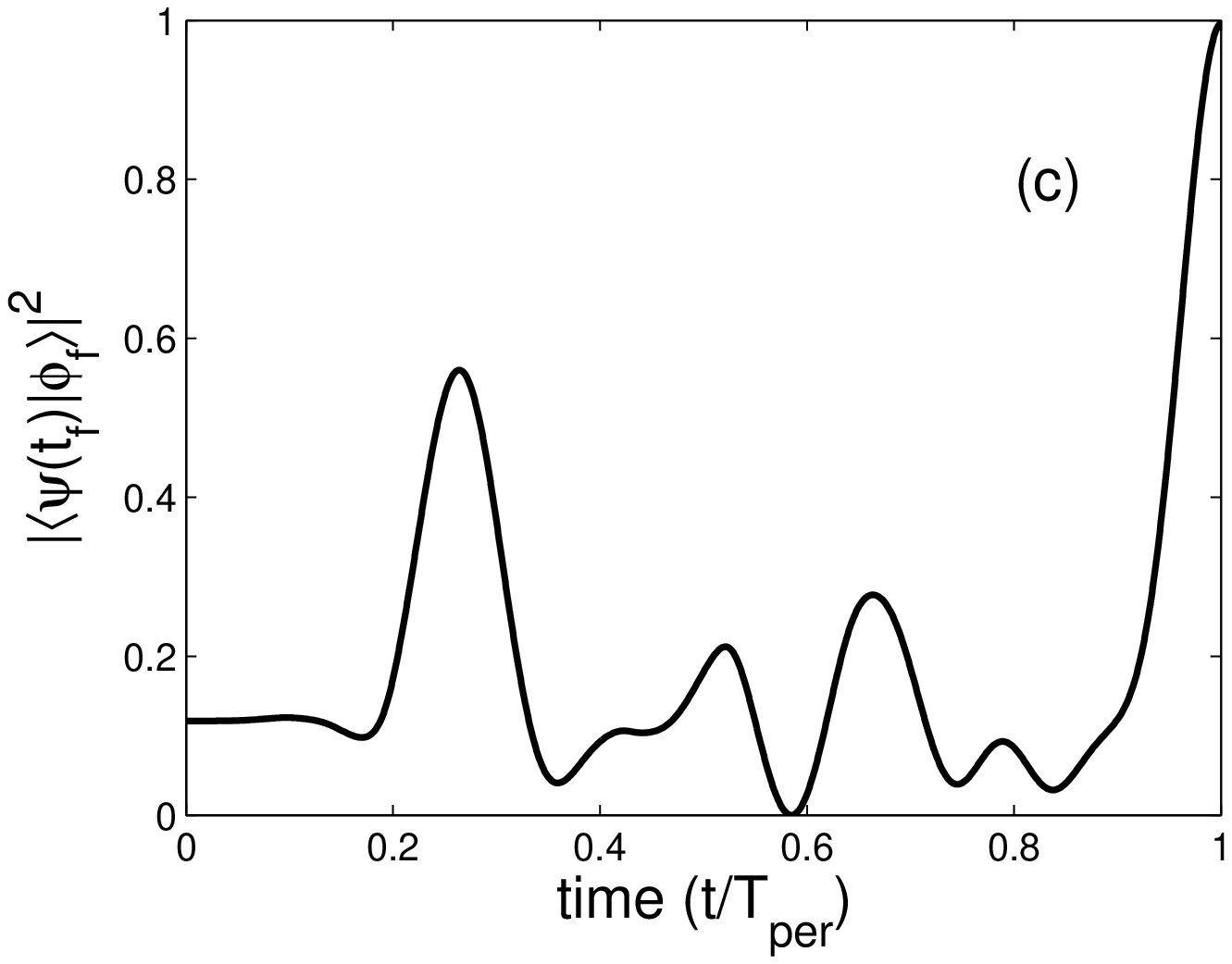}
\caption{\label{fig1} Plot as a function of the adimensional time
$t/T_{per}$ of (a) the optimal field (solid line) and the initial
trial field (dashed line), (b) the expectation value
$\langle\cos\theta\rangle$ and (c) the projection onto the target
state $|\phi_f\rangle$. The abbreviation a.u. corresponds to
atomic units.}
\end{figure}
\begin{figure}
\includegraphics[height=1.75in]{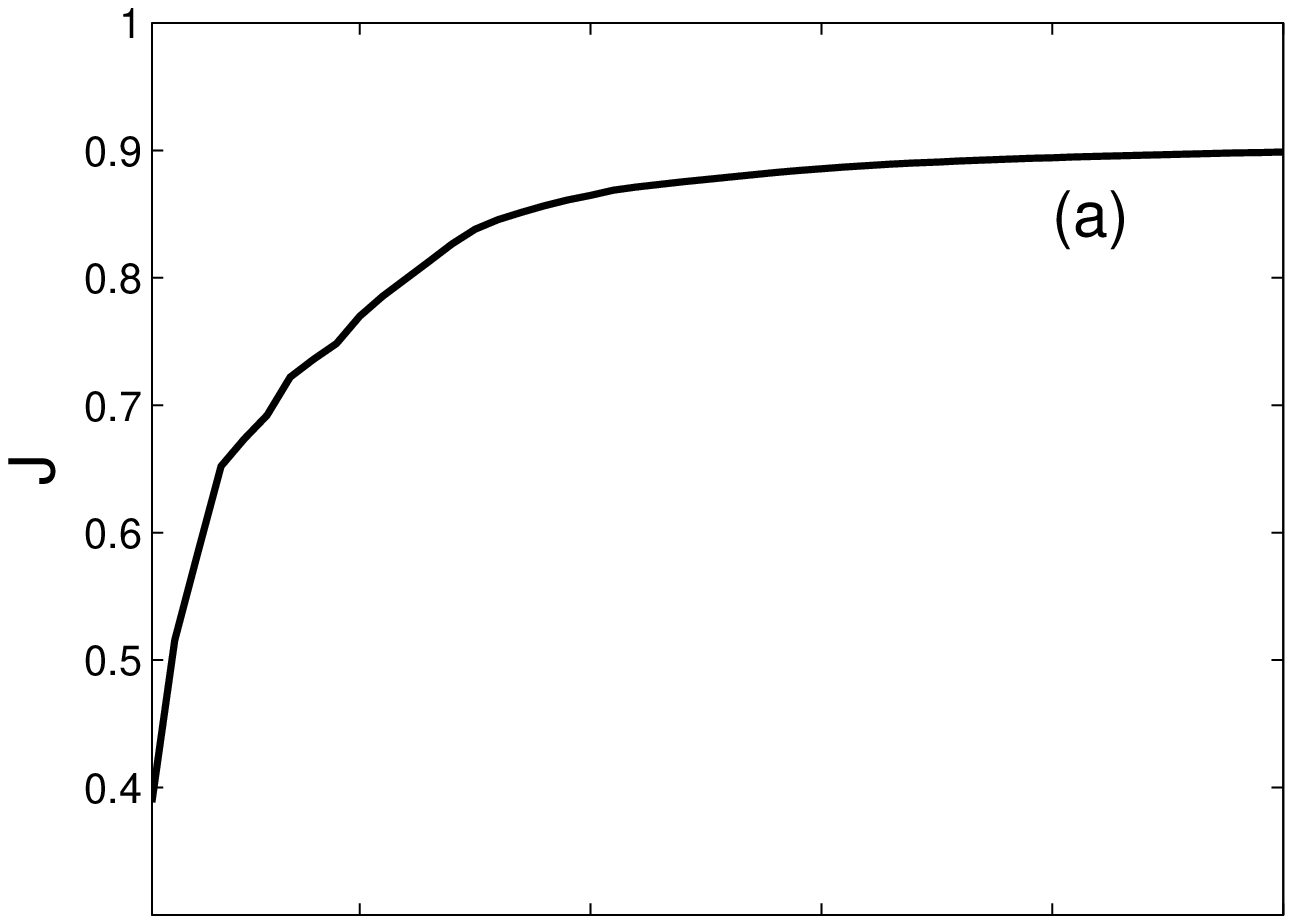}
\includegraphics[height=1.75in]{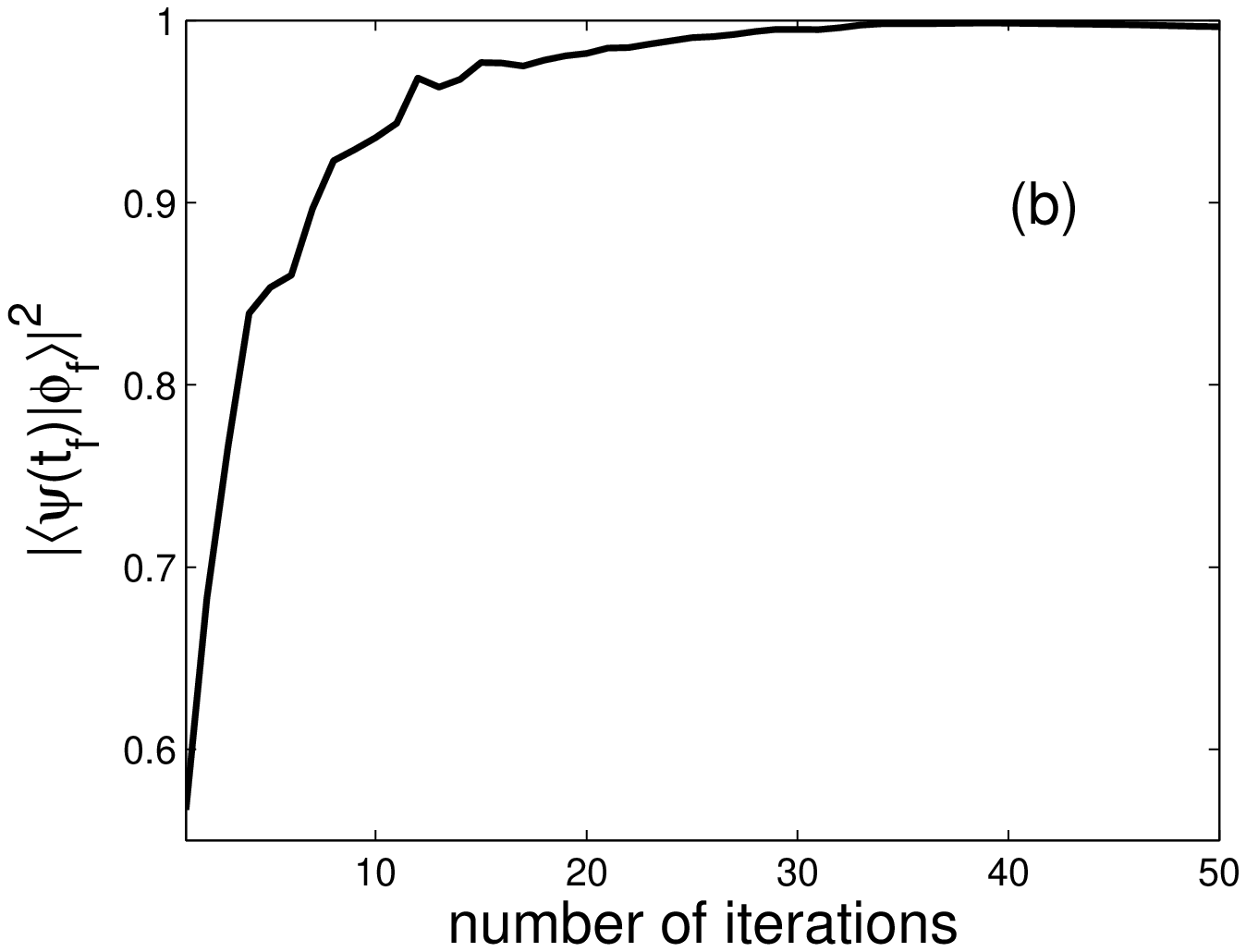}
\caption{\label{fig2} Plot as a function of the number of
iterations of (a) the adimensional cost $J$  defined by Eq.
(\ref{eq12}) for $n=1$ (a cost quadratic in the field) and (b) the
projection onto the target state at time $t_f$.}
\end{figure}
\begin{figure}
\includegraphics[height=1.75in]{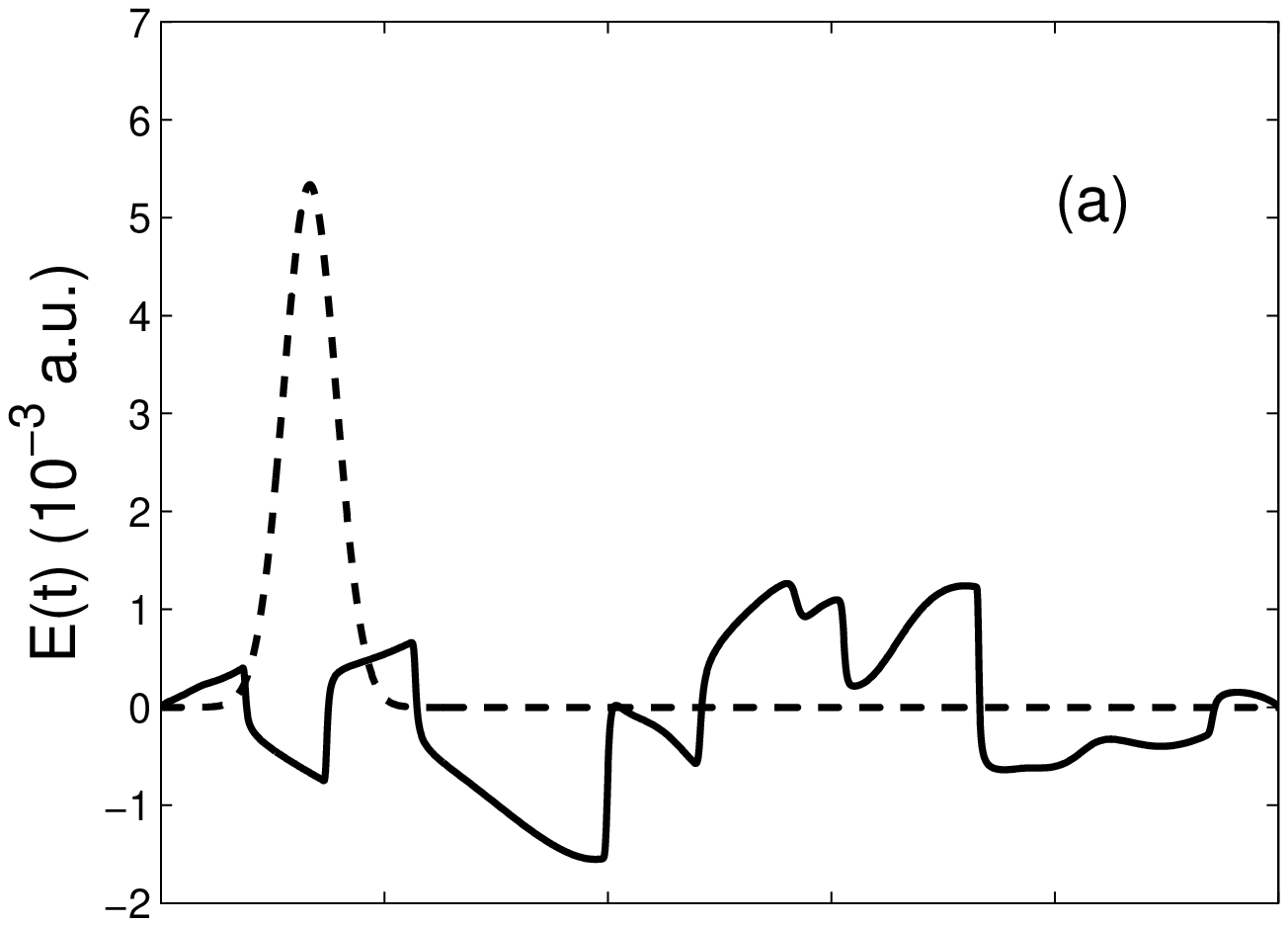}
\includegraphics[height=1.75in]{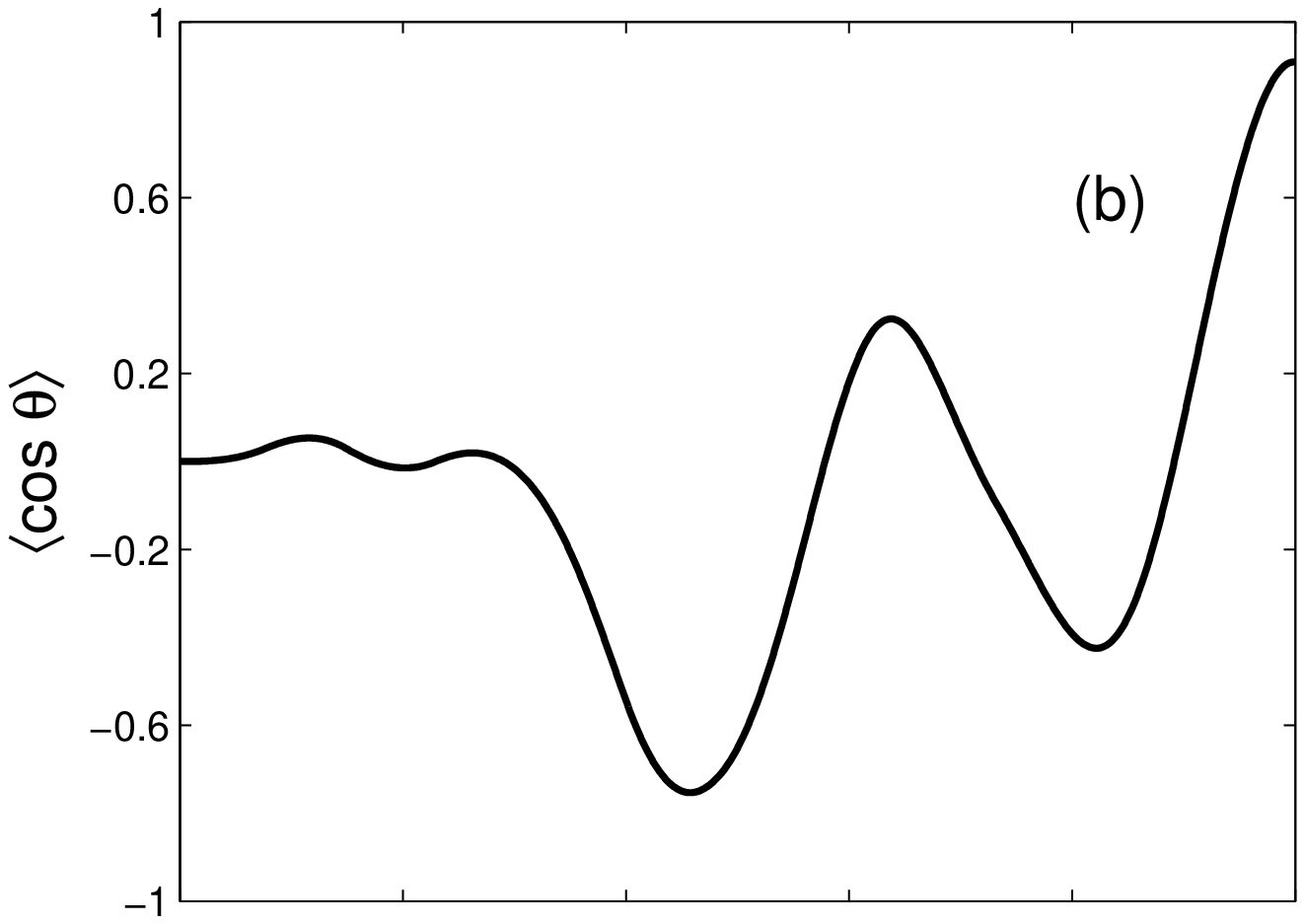}
\includegraphics[height=1.75in]{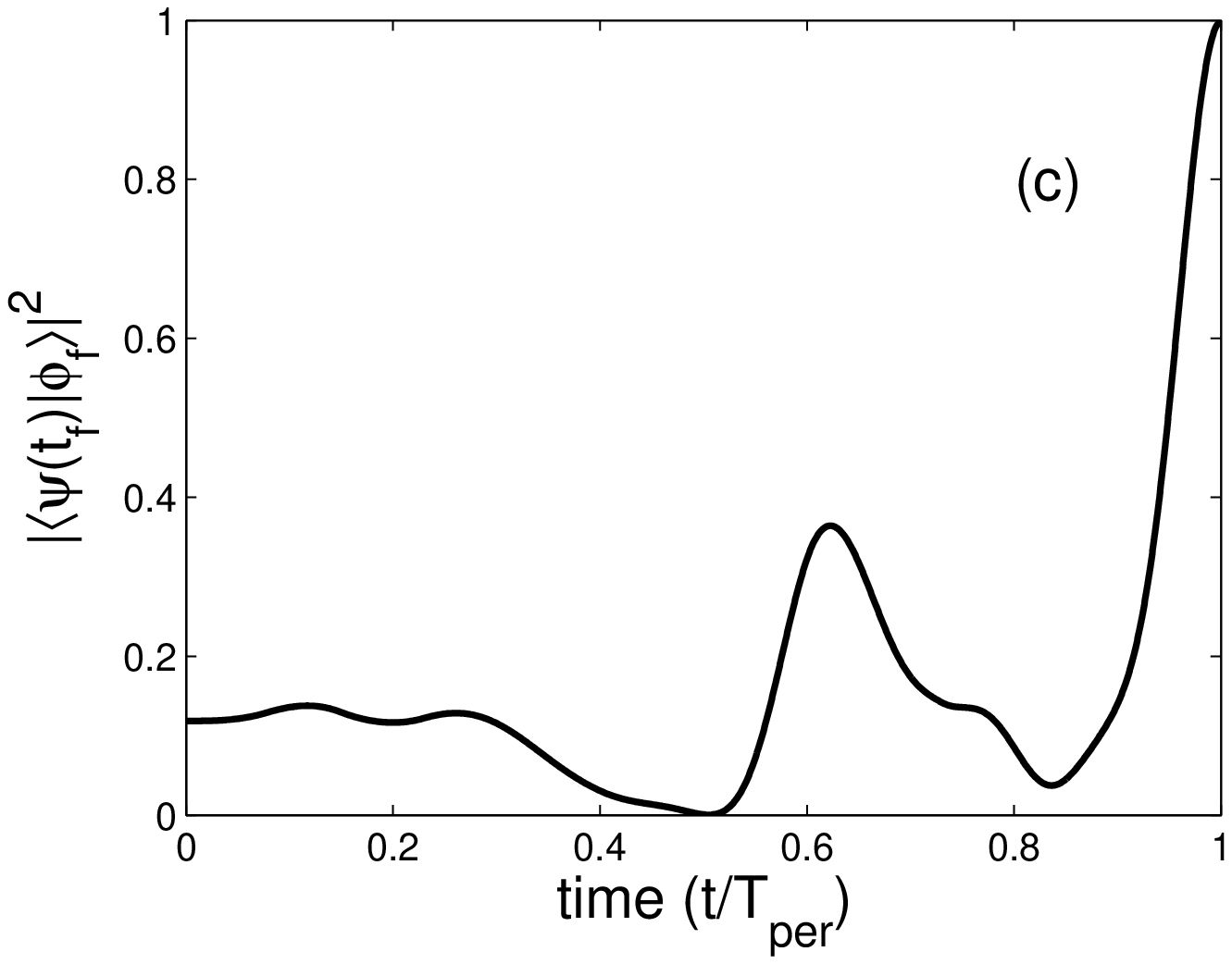}
\caption{\label{fig3} Same as Fig. \ref{fig1} but for $n=2$, i.e.,
a cost quartic in the field.}
\end{figure}
\begin{figure}
\includegraphics[height=1.75in]{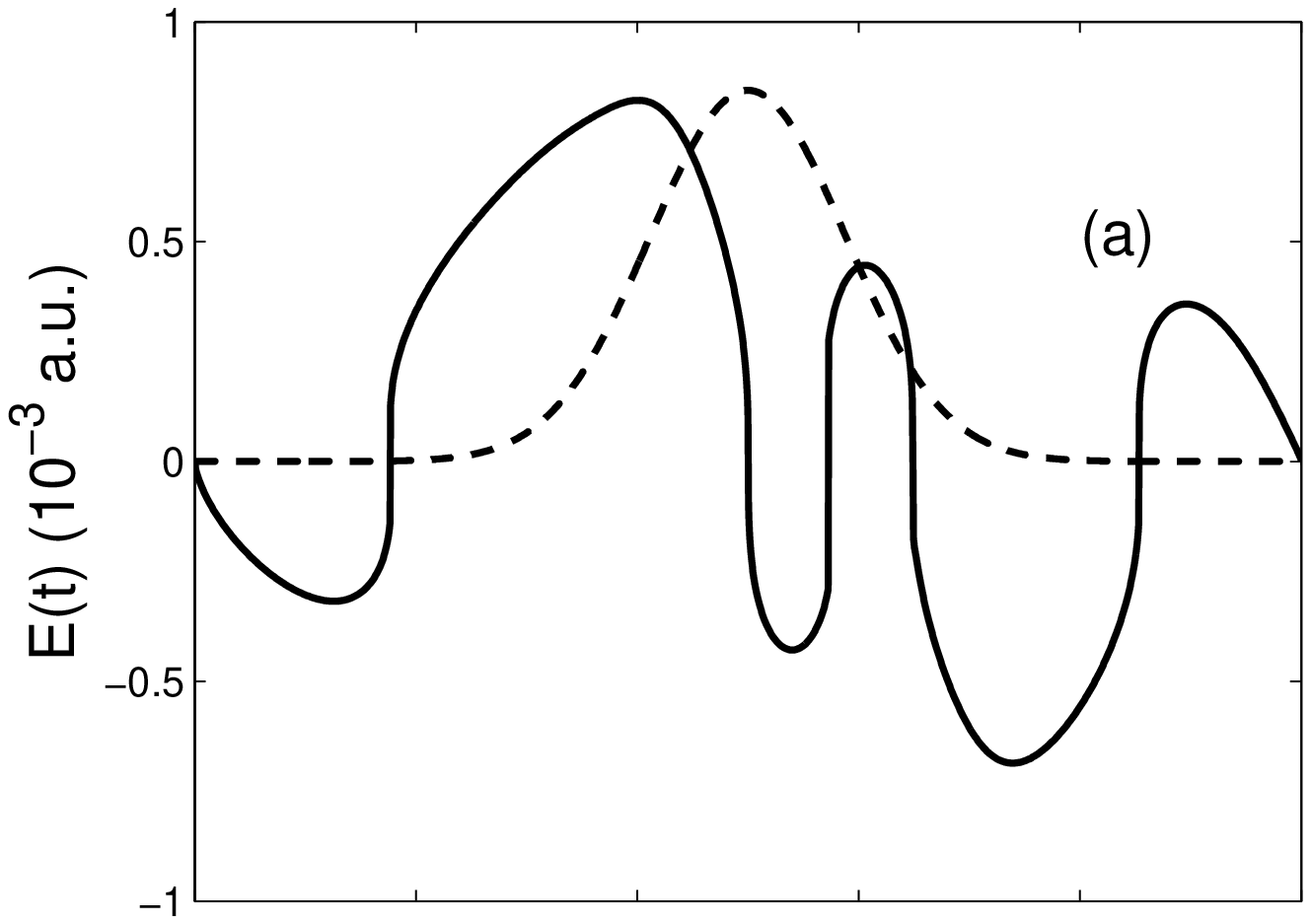}
\includegraphics[height=1.75in]{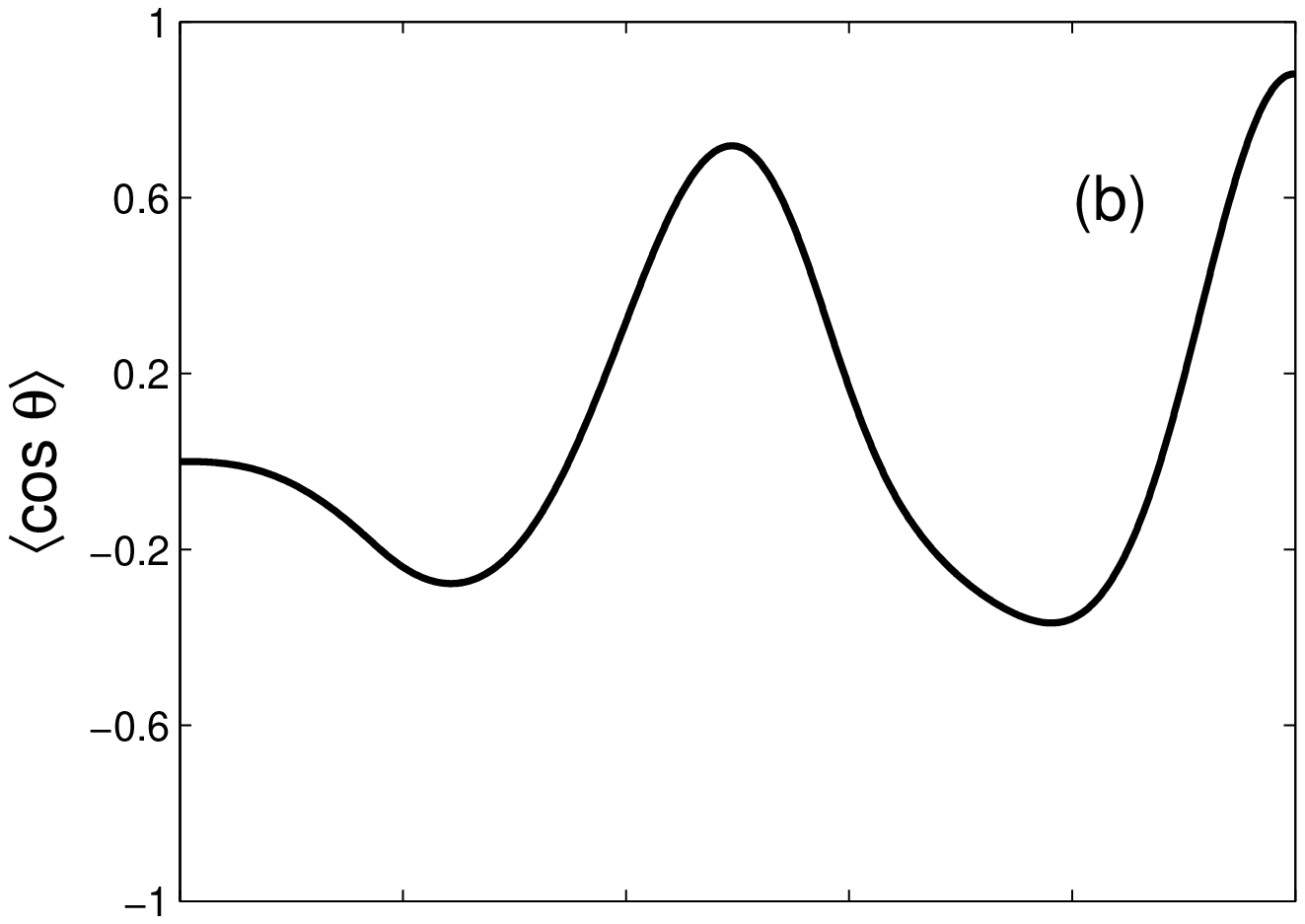}
\includegraphics[height=1.75in]{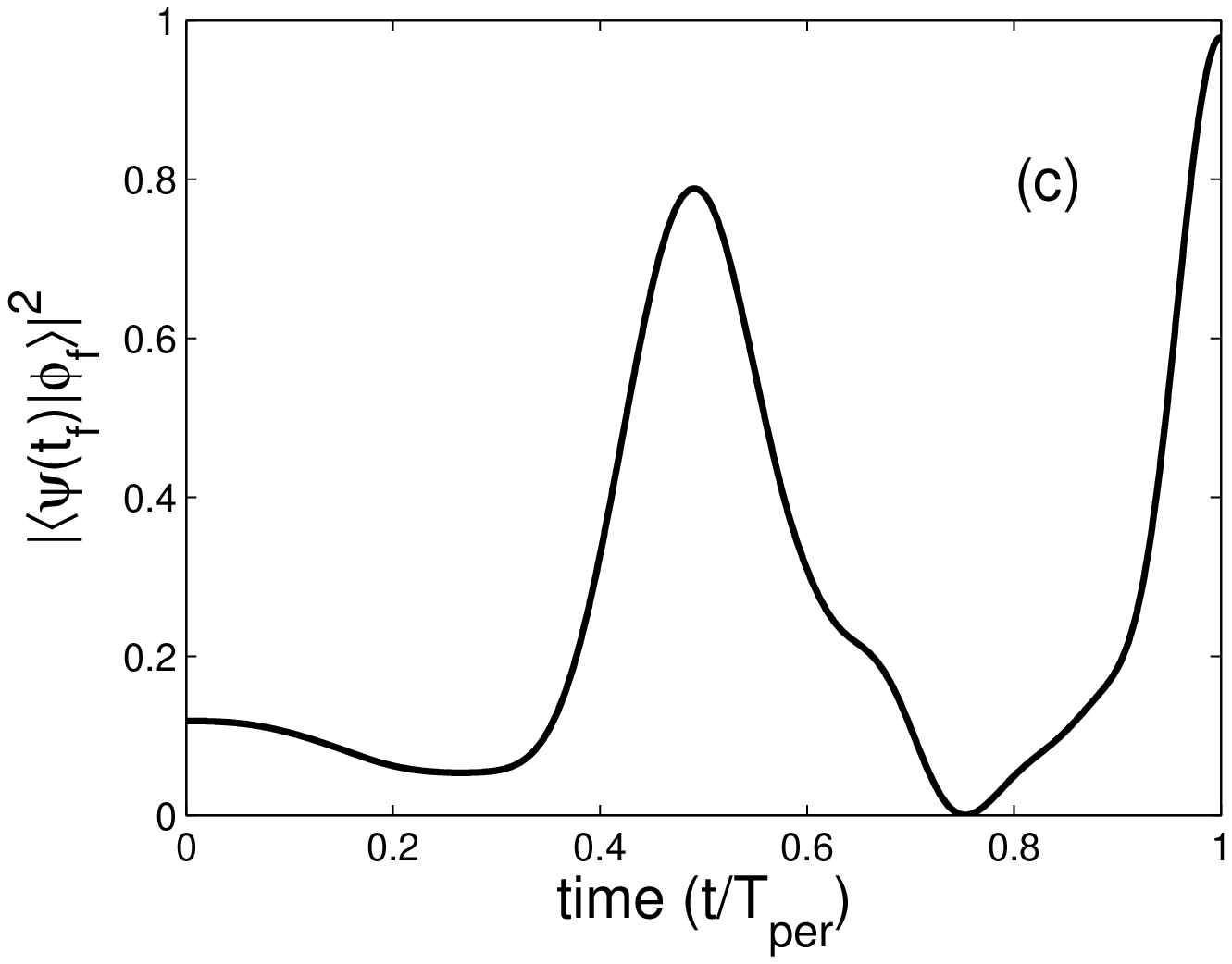}
\caption{\label{fig4} Same as Fig. \ref{fig1} but for the
algorithm I and $n=2$.}
\end{figure}
The results of the computations are presented in Figs. \ref{fig1},
\ref{fig2}, \ref{fig3} and \ref{fig4}. We have used the simplified
algorithms by assuming that $E_k=\tilde{E}_k$. Figures \ref{fig1},
\ref{fig3} and \ref{fig4} correspond respectively to the algorithm
II for $n=1$, the algorithm II for $n=2$ and the algorithm I for
$n=2$. For the algorithm II, we can choose $n=1$ since we have
checked that for this value Eqs. (\ref{eq25}) and (\ref{eq26})
always have a real solution. Note that this latter observation
depends on the values of $\lambda$ and $\eta$ considered.
Numerical values are taken to be $\lambda=0.05,\eta=1$ for Fig.
\ref{fig1} ($\lambda$ corresponds to the maximum value of
$\lambda(t)$), $\lambda=6.05\times 10^4,\eta=1$ for Fig.
\ref{fig3} and $\lambda=12\times 10^5$ for Fig. \ref{fig4}. The
difference in the values of $\lambda$ is due to the form of the
cost which is either quadratic or quartic in the field. We have
checked that the value of $\eta$ is not relevant even if the value
of $\lambda$ has to be adjusted with respect to the one of $\eta$.
The trial fields are displayed in Fig. \ref{fig1}a, \ref{fig3}a
and \ref{fig4}a. The trial field is a gaussian pulse of intensity
of the order of 1 $\textrm{TW}/\textrm{cm}^2$. In order to obtain
realistic electric fields, the value of $\lambda$ has been chosen
so that the energy of the optimal field be lower than two times
the energy of the gaussian pulse. Note also that for $E\simeq
5.10^{-3}\textrm{a.u.}$ (which corresponds to the typical
amplitude of the optimal field), we have $\mu_0\simeq \alpha_\perp
E$ which shows that the polarizability terms are not negligible in
the dynamics. In each case, very good results are obtained with a
final projection $|\langle\psi(t_f)|\phi_f\rangle|^2$ larger than
0.99 except for the algorithm I where
$|\langle\psi(t_f)|\phi_f\rangle|^2$ is of the order of 0.98.
Figure \ref{fig2} illustrates the convergence properties of the
algorithm II which are satisfactory since after 30 iterations, we
obtain a projection close to 0.98. A similar behavior has been
observed in the other cases. A comparison of Figs. \ref{fig1}a,
\ref{fig3}a and \ref{fig4}a shows that the optimal field for $n=2$
has sharper variations than for $n=1$ for both algorithms. We have
also observed that these sharper variations can induce numerical
instabilities and high frequency oscillations in the optimal
field. This point is discussed in Sec. \ref{sec3c} where we show
how to remove the parasite oscillations with a band-pass filter.
In practice, it has been found that small values of the exponent
$n$ generally produce smoother optimal fields.
\subsection{Non-resonant two-color laser fields}\label{sec3b}
We continue to consider a zero rotational temperature but we
assume now that the molecule interacts with a non-resonant
two-color laser field \cite{friedrich,tehini} of the form
\begin{equation}\label{eqav1}
E(t)=E_1(t)\cos(\omega t)+E_2(t)\cos(2\omega t).
\end{equation}
After averaging over the rapid oscillations of the field, the
Hamiltonian $\hat{H}$ of the system becomes
\begin{eqnarray}\label{eqav2}
&
&\hat{H}=B\hat{J}^2-\frac{1}{4}[(\alpha_\parallel-\alpha_\perp)\cos^2\theta+\alpha_\perp](E_1(t)^2+E_2(t)^2)\nonumber
\\& &
-\frac{1}{8}[(\beta_\parallel-3\beta_\perp)\cos^3\theta+3\beta_\perp\cos\theta]E_1(t)^2E_2(t).
\end{eqnarray}
The interest of this model is due to the absence of linear term in
the interaction which enhances the difficulty of the control. Two
cases can be considered according to the respective values of
$E_1$ and $E_2$. If $E_1=E_2$, we can use the standard algorithm
presented in Sec. \ref{sec2c} whereas for $E_1\neq E_2$, the
algorithm has to be slightly generalized. These two problems are
respectively analyzed in Secs. \ref{sec3b1} and \ref{sec3b2}.
\subsubsection{The case $E_1\neq E_2$}\label{sec3b1}
We first generalized the algorithm of \ref{sec2c} to the case of
two control fields. We assume that the cost is quadratic in the
field.

We introduce the augmented cost $\bar{J}$ and we determine the
critical points with respect to $E_1$ and $E_2$. We have:
\begin{eqnarray}\label{eqav3}
& &\bar{J}=|\langle \phi_f|\psi(t_f)\rangle|^2-\int_0^{t_f}\lambda
[E_1(t)^2+E_2(t)^2]dt\nonumber\\
& &
-2\Im[\langle\psi(t_f)|\phi_f\rangle\int_0^{t_f}\langle\chi(t)|(i\frac{\partial}{\partial
t}-\hat{H})|\psi(t)\rangle dt],
\end{eqnarray}
and we compute the variational derivatives $\delta\bar{J}/\delta
E_1$ and $\delta\bar{J}/\delta E_2$ which are equal to zero for a
critical point. We then obtained the following system of equations
\begin{eqnarray}\label{eqav4}
\begin{array}{ll}
\lambda E_1+2\tilde{\alpha}E_1+2\tilde{\beta}E_1E_2=0 \\
\lambda E_2+2\tilde{\alpha}E_2+\tilde{\beta}E_1^2=0
\end{array}
\end{eqnarray}
which are satisfied by the optimal fields $E_1$ and $E_2$. We have
used in Eqs. (\ref{eqav4}) the notations
\begin{eqnarray}\label{eqav5}
\begin{array}{ll}
\tilde{\alpha}=2\Im[\langle\psi(t)|\chi(t)\rangle\langle
\chi(t)|\frac{1}{4}((\alpha_\parallel-\alpha_\perp)\cos^2\theta+\alpha_\perp)|\psi(t)\rangle]\\
\tilde{\beta}=2\Im[\langle\psi(t)|\chi(t)\rangle\langle
\chi(t)|\frac{1}{8}[(\beta_\parallel-3\beta_\perp)\cos^3\theta+3\beta_\perp\cos\theta]|\psi(t)\rangle].
\end{array}
\end{eqnarray}
We solve the optimal equations by a monotonically convergent
algorithm. The proof of monotonicity follows closely the lines of
proof in Sec. \ref{sec2c}. We use the same notations, with for
instance $E_{1,k}$ the field $E_1$ at iteration $k$. To simplify
the computations, we take equal the fields $E_k$ and $\tilde{E}_k$
for the forward and the backward propagations. We compute $\Delta
J=J_{k+1}-J_k$. We obtain the following expressions for the
polynomials $\mathcal{P}_1$ and $\mathcal{P}_2$:
\begin{eqnarray}\label{eqav6}
&
&\mathcal{P}_1=(E_{1,k+1}-E_{1,k})[(E_{1,k+1}+E_{1,k})\alpha_{k,k+1}+
\\& &(E_{1,k+1}+E_{1,k})E_{2,k+1}\beta_{k,k+1}-\lambda (E_{1,k+1}+E_{1,k})]\nonumber
\end{eqnarray}
and
\begin{eqnarray}\label{eqav7}
&
&\mathcal{P}_2=(E_{2,k+1}-E_{2,k})[(E_{2,k+1}+E_{2,k})\alpha_{k,k+1}+
\\& & E_{1,k}^2\beta_{k,k+1}-\lambda (E_{2,k+1}+E_{2,k})]\nonumber
\end{eqnarray}
where
\begin{eqnarray}\label{eqav8}
\begin{array}{ll}
\alpha_{k,k+1}=\Im[\langle\psi_{k+1}(t)|\chi_k(t)\rangle\langle
\chi_k(t)|\frac{1}{4}((\alpha_\parallel-\alpha_\perp)+\alpha_\perp)\cos^2\theta|\psi_{k+1}(t)\rangle]\\
\beta_{k,k+1}=\Im[\langle\psi_{k+1}(t)|\chi_k(t)\rangle\langle
\chi_k(t)|\frac{1}{8}[(\beta_\parallel-3\beta_\perp)\cos^3\theta+3\beta_\perp\cos\theta]|\psi_{k+1}(t)\rangle].
\end{array}
\end{eqnarray}
$\mathcal{P}_1$ and $\mathcal{P}_2$ are respectively viewed as
polynomials in $E_{1,k+1}$ and $E_{2,k+1}$. We first use
$\mathcal{P}_2$ to determine the field $E_{2,k+1}$ by the
algorithm I or II  and then using this solution, we compute
$E_{1,k+1}$ from $\mathcal{P}_1$. We also check that if the
algorithm converges then the solutions given by the algorithm
correspond to the extremal solutions defined by Eqs.
(\ref{eqav4}). This can be done by replacing $E_{1,k+1}$ and
$E_{1,k}$ by $E_1$ and $E_{2,k+1}$ and $E_{2,k}$ by $E_2$. The
optimal fields are then zeros of the derivatives of
$\mathcal{P}_1$ and $\mathcal{P}_2$ with respect to $E_{1,k+1}$
and $E_{2,k+1}$.
\begin{figure}
\includegraphics[height=1.75in]{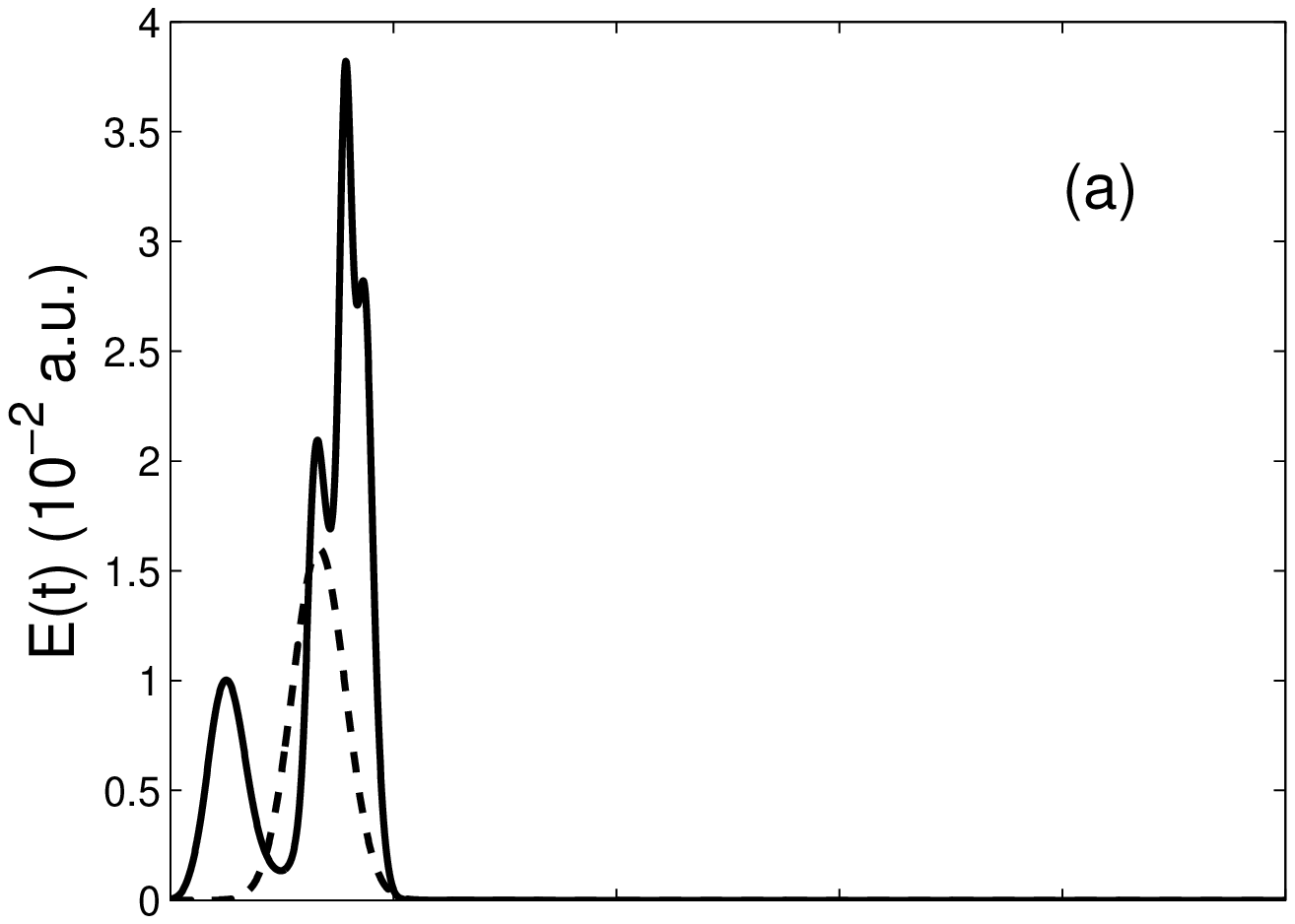}
\includegraphics[height=1.75in]{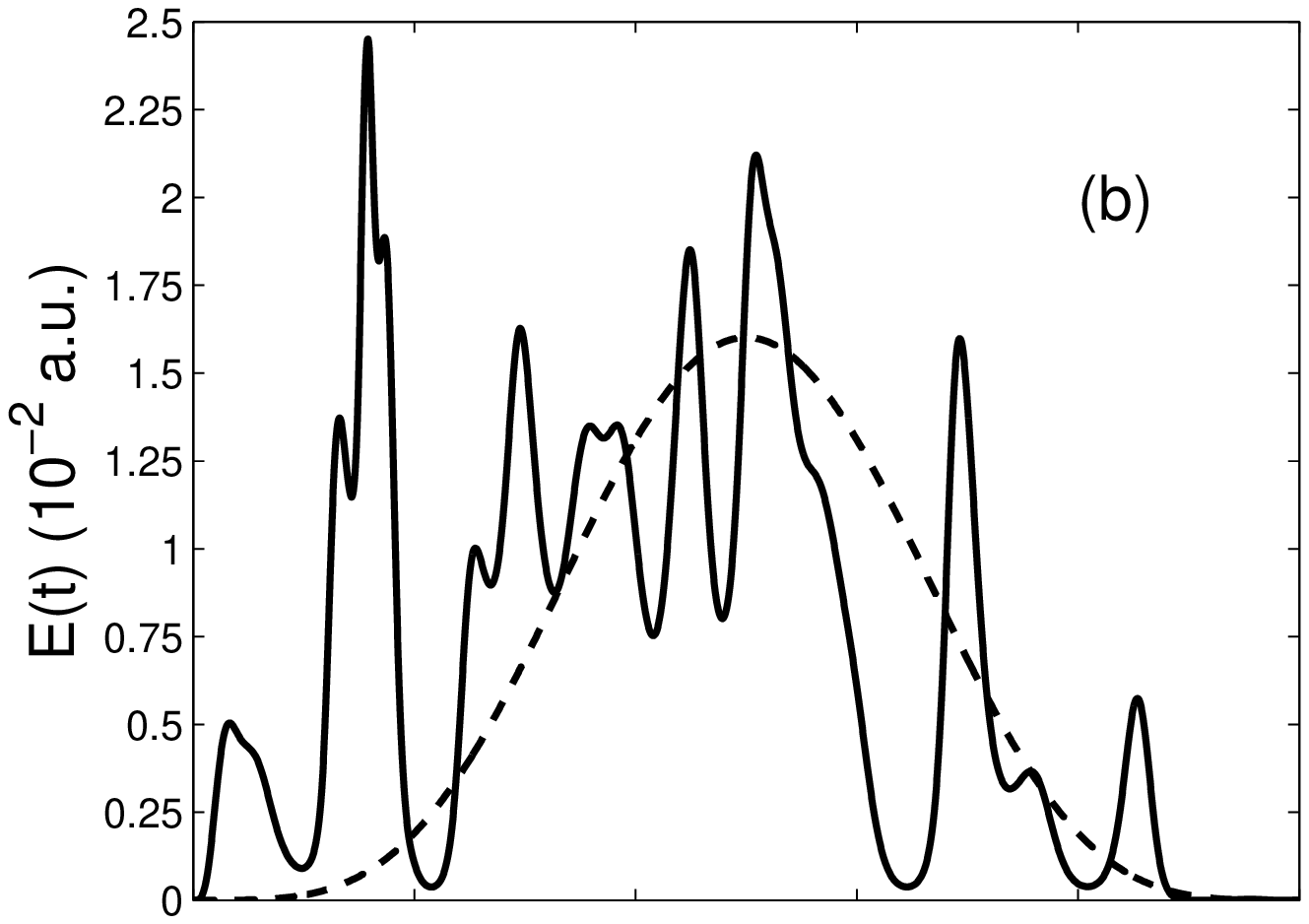}
\includegraphics[height=1.75in]{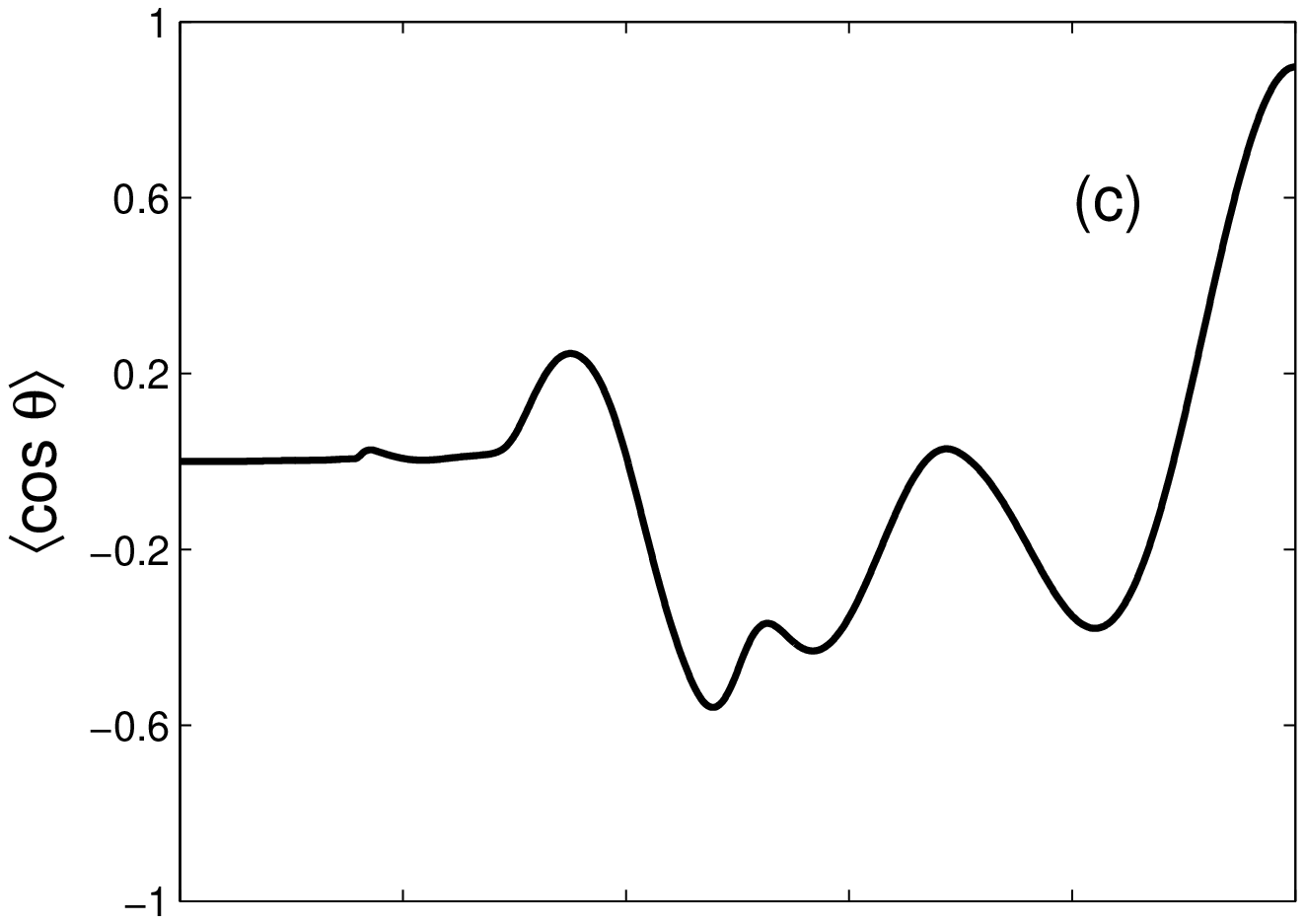}
\includegraphics[height=1.75in]{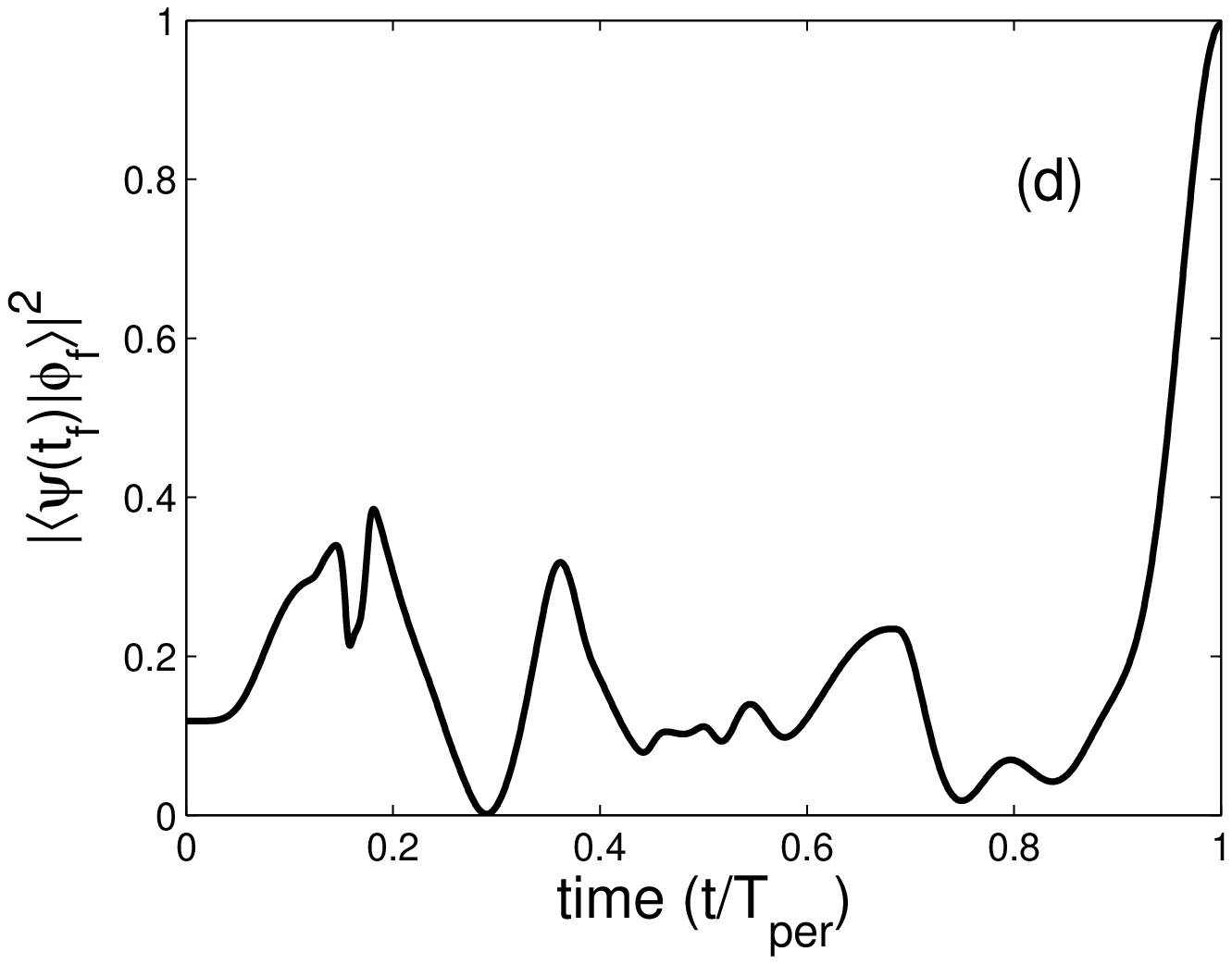}
\caption{\label{fig5} Same as Fig. \ref{fig1} but for two-color
laser fields with $E_1\neq E_2$. The panels (a) and (b) correspond
respectively to the fields $E_1$ and $E_2$.}
\end{figure}
Figure \ref{fig5} displays the results we have obtained with the
algorithm II for $n=2$ and $\lambda=1$.We have chosen two
different trial fields in order to generate two different optimal
fields $E_1$ and $E_2$. With the same trial field for the fields
$E_1$ and $E_2$, the algorithm leads to two solutions which are
very close to each other. Larger values of electric fields have
been used due to the absence of linear interaction term in the
Hamiltonian.

A remarkable characteristic of the optimal fields is the fact that
$E_1$ vanishes for $t>0.2\times T_{per}$. This means that the
dissymmetry producing the orientation (dissymmetry due to the term
in $E_1^2E_2$ in the Hamiltonian) only acts during this duration.
This provides a non-intuitive and new method to produce
orientation using a long laser field $E_2$ and a short laser field
$E_1$.
\subsubsection{The case $E_1=E_2$}\label{sec3b2}
\begin{figure}
\includegraphics[height=1.75in]{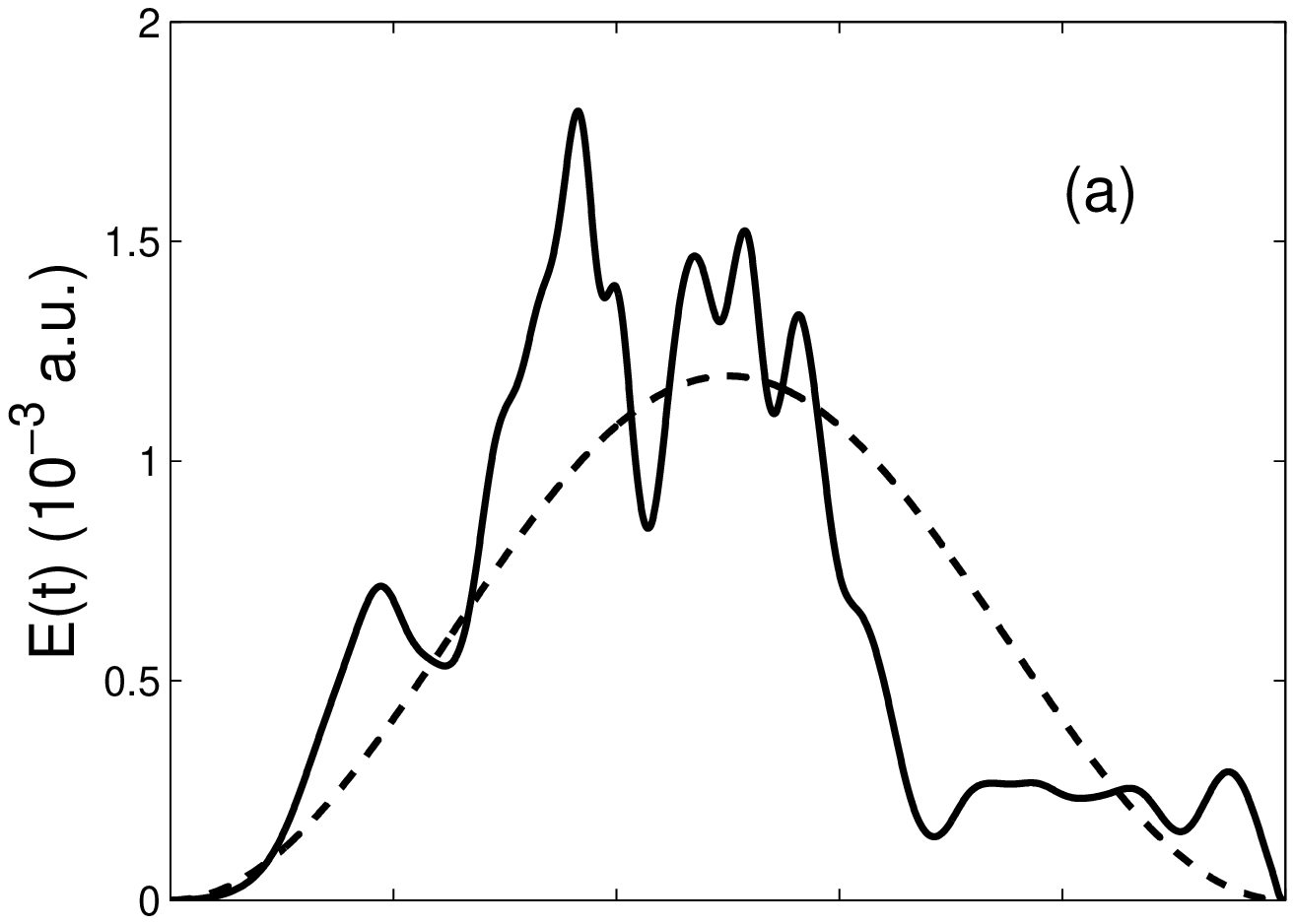}
\includegraphics[height=1.75in]{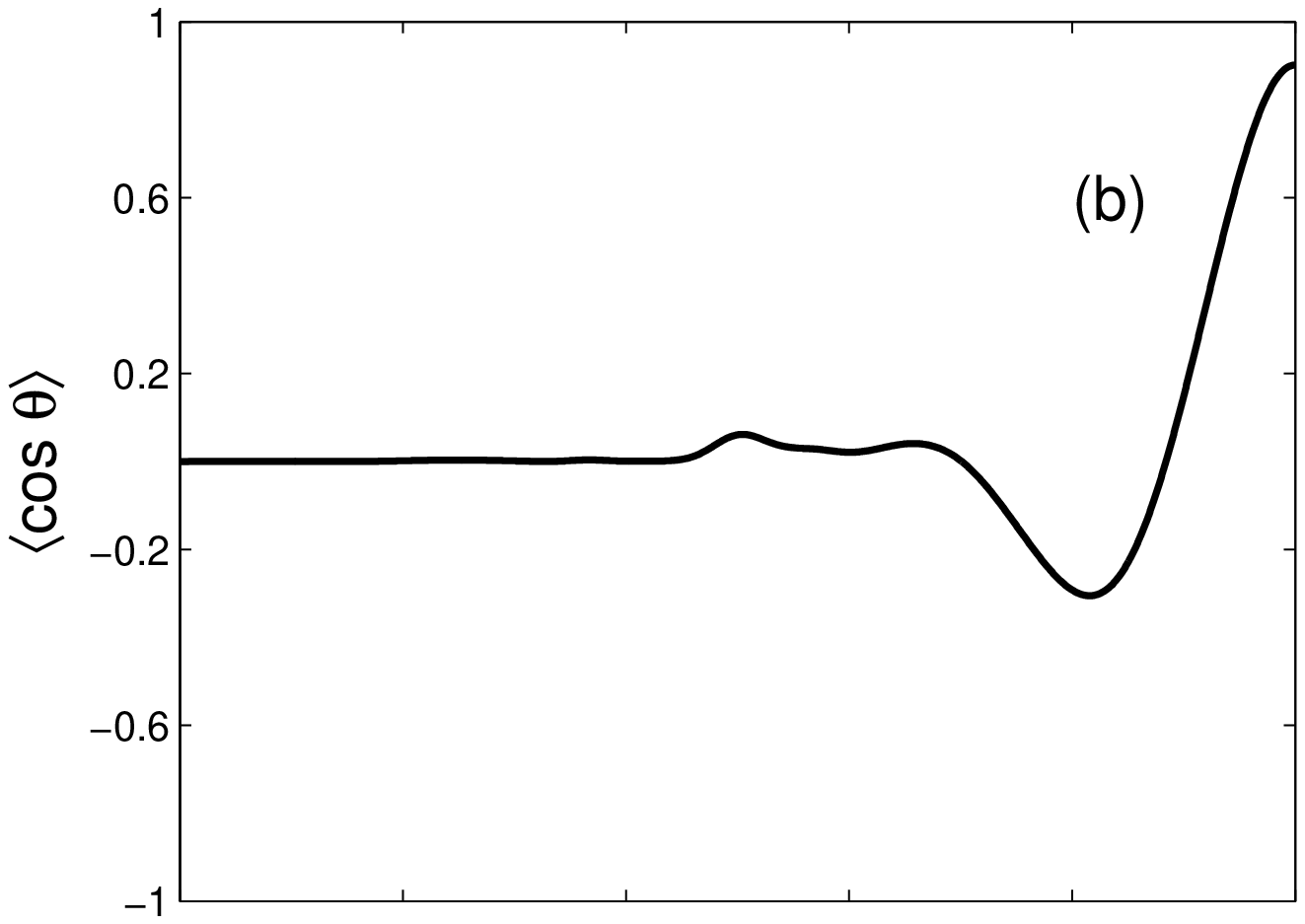}
\includegraphics[height=1.75in]{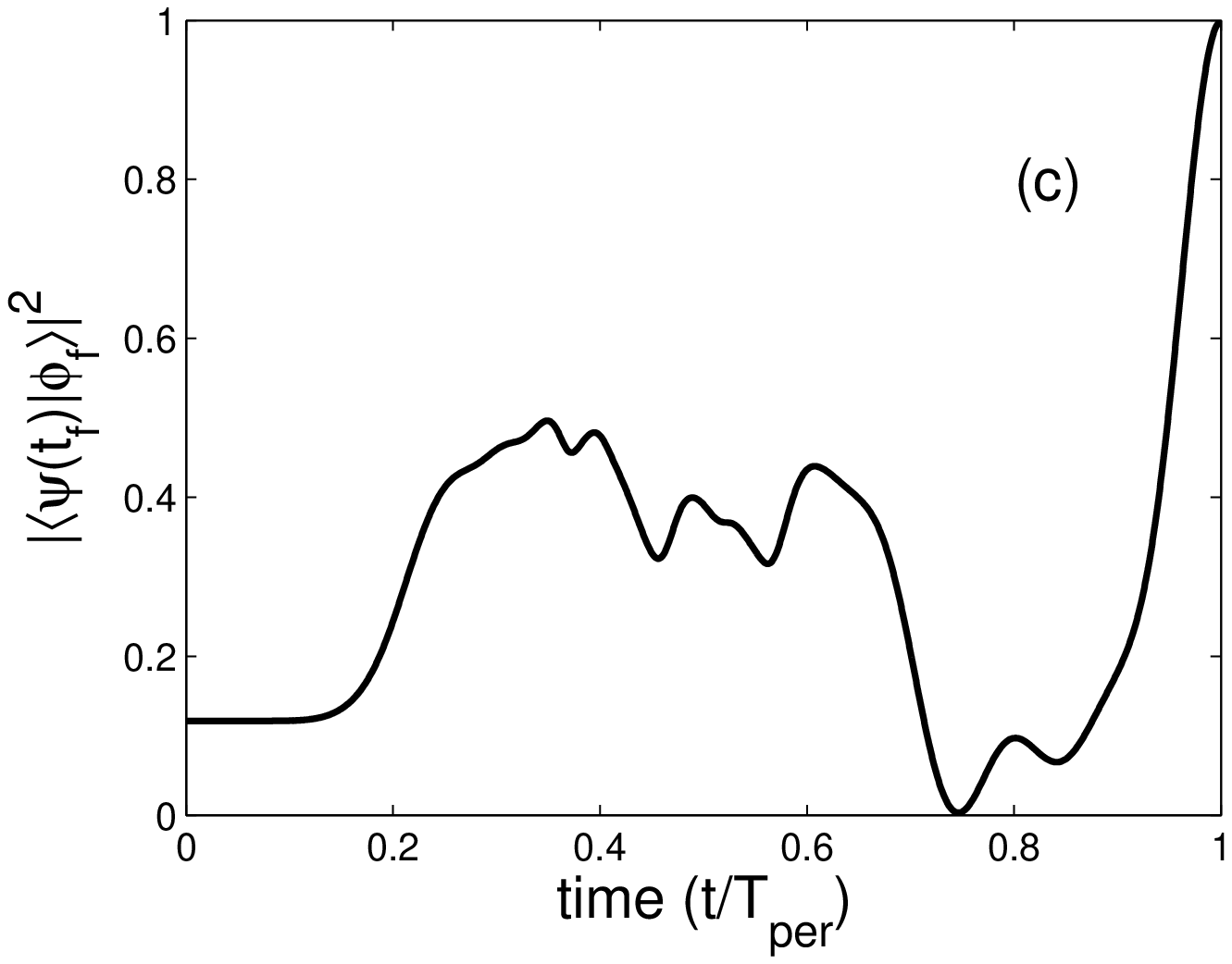}
\caption{\label{fig6} Same as Fig. \ref{fig1} but for the
averaging case with $E_1=E_2$.}
\end{figure}

We use the monotonic algorithm II proposed in Sec. \ref{sec2}.
Figure \ref{fig6} illustrates the different results. They have
been obtained for $n=2$ and $\lambda=5$. The trial field is a
gaussian pulse whose duration corresponds to the rotational
period. From the equations of the algorithm, it is straightforward
to see that the algorithm cannot generate an optimal field
different from zero at time $t$ if the trial field is zero at that
time. This is simply due to the absence of linear interaction term
in the Hamiltonian. The algorithm only modifies the envelope of
the trial field whose choice is therefore crucial.
\subsection{Analysis of the Fourier spectrum}\label{sec3c}
We analyze in this section the Fourier transforms of the optimal
solutions. Our goal is to show that optimal solutions determined
in Sec. \ref{sec2a} can be well-approximated by solutions that
could be implemented experimentally. We consider experiments
coupled with genetic algorithms optimizing the phase and the
amplitude of the Fourier transform of a finite number of frequency
components. The discretization is done over a frequency interval
chosen with respect to the quantum transition frequencies involved
in the control (see below). Note that we do not take into account,
in this paper, technology constraints for the choice of this
frequency interval. Standard pulse shapers usually work with
optical frequencies of the order of 800 nm. With such a
technology, only non-resonant laser fields of Sec. \ref{sec3b}
could be experimentally implemented.

Following \cite{hertz1,hertz2}, we assume that the solution
obtained by genetic algorithms is a piecewise constant function in
frequency both in amplitude and in phase. We have chosen 640
frequencies or less to discretize the optimal field. By an inverse
Fourier transform, we then determine a new time-dependent electric
field.
\begin{figure}
\includegraphics[height=1.75in]{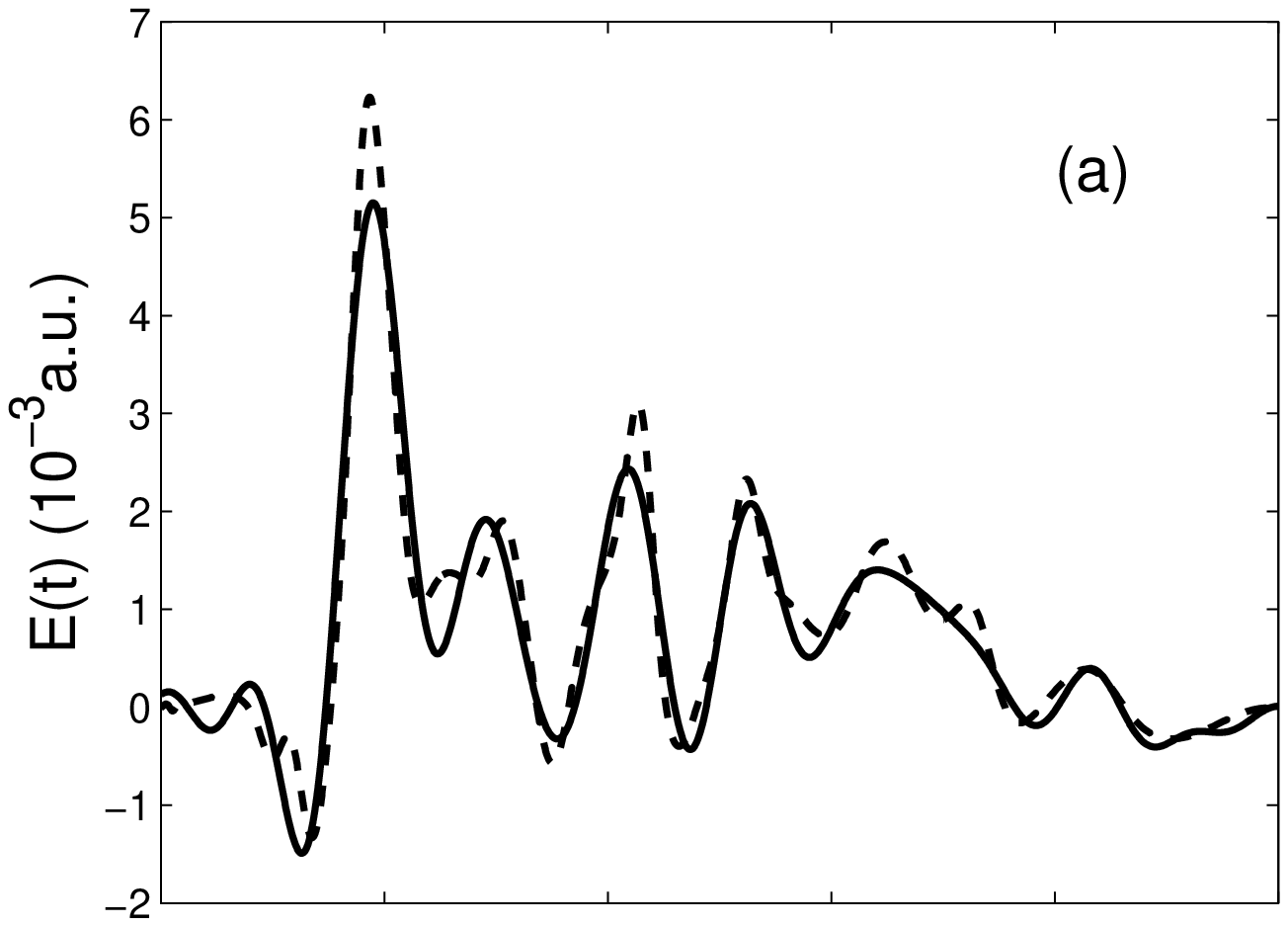}
\includegraphics[height=1.75in]{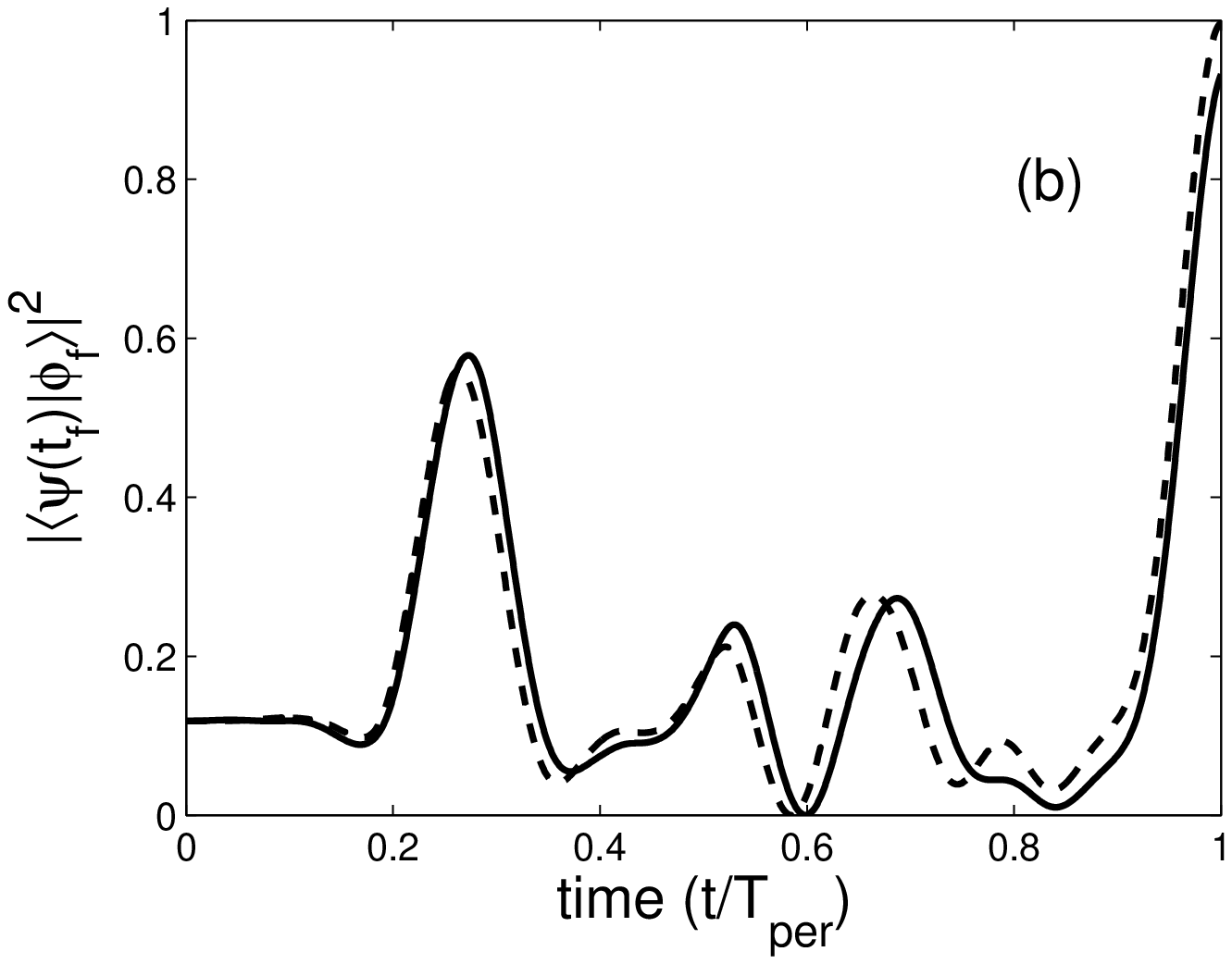}
\caption{\label{fig7} Plot as a function of time $t/T_{per}$ of
(a) the optimal (dashed line) and of the approximate fields (solid
line) and (b) the projection onto the target state
$|\phi_f\rangle$.}
\end{figure}
Figure \ref{fig7} presents the results obtained with the optimal
pulse of Fig. \ref{fig1}. Using only 128 frequencies, we show that
the final projection obtained by the optimal pulse and its
approximation are very close to each other. For 256 frequencies,
the difference is negligible and cannot be distinguished at the
resolution of the plots.
\begin{figure}
\includegraphics[height=1.75in]{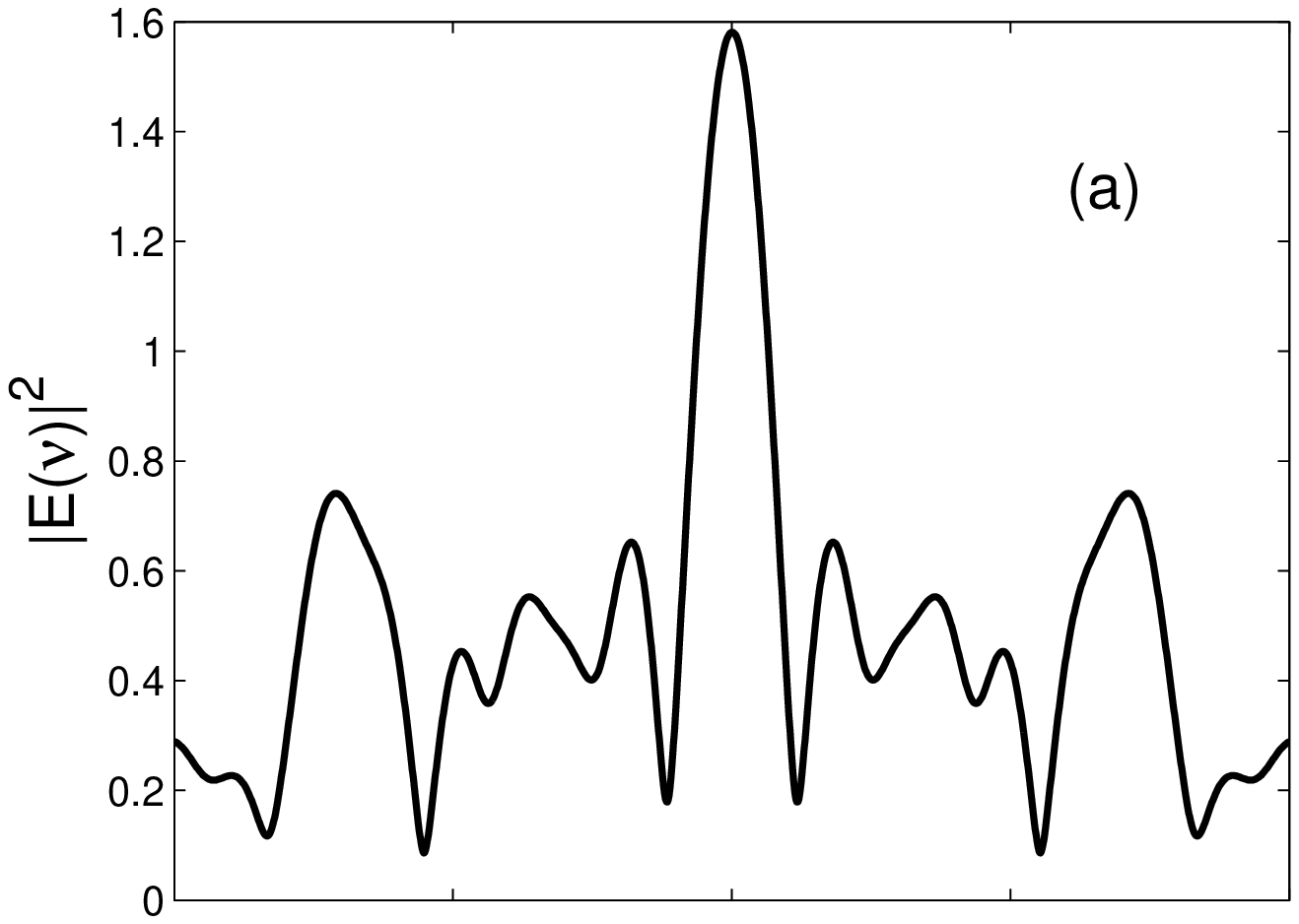}
\includegraphics[height=1.75in]{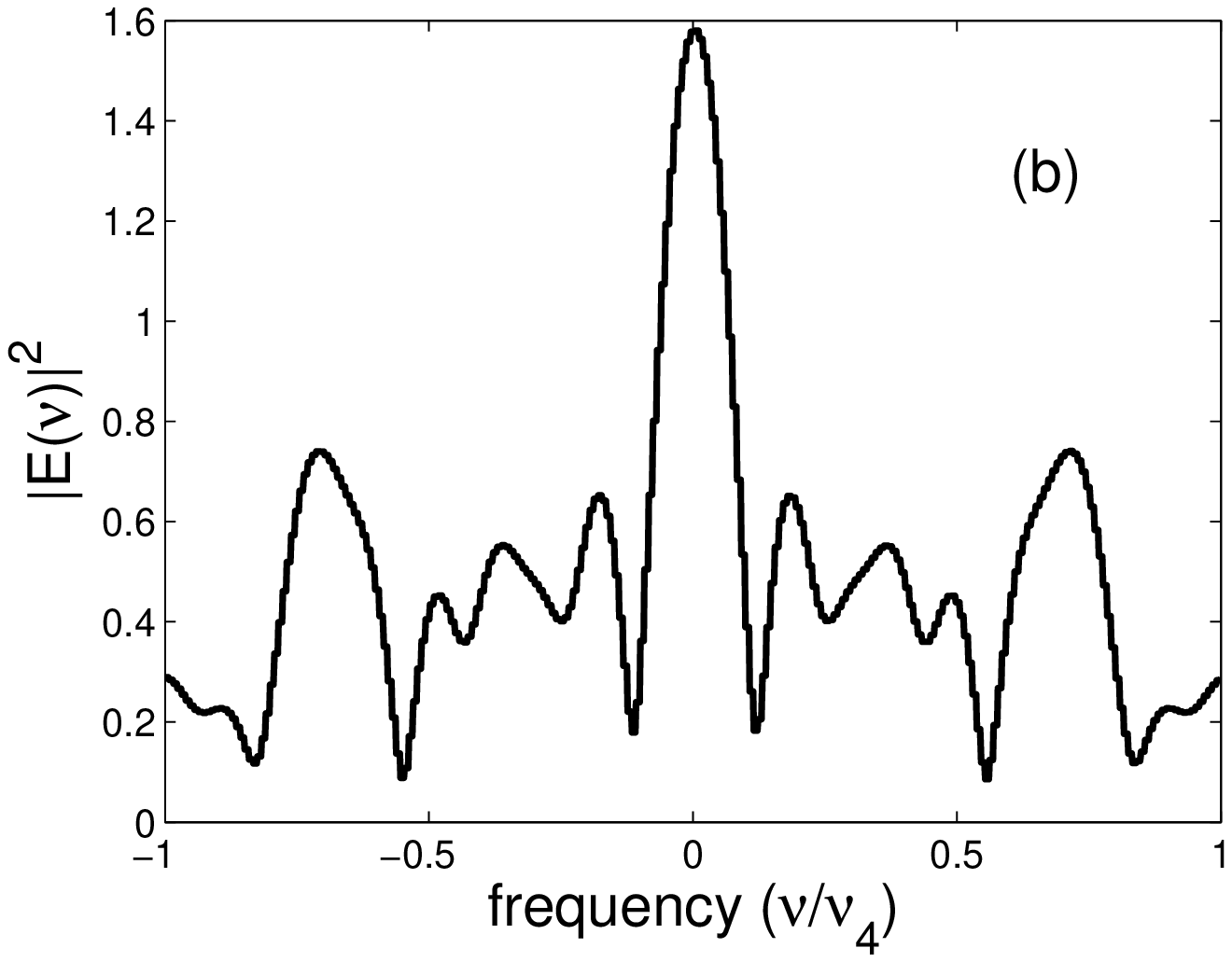}
\caption{\label{fig8} Plot as a function of the frequency $\nu$ of
the module square of the Fourier transform of the optimal field
(a) and of its piecewise constant approximation (b). The Fourier
transform has been discretized over the interval
$[-\nu_4,\nu_4]$.}
\end{figure}
Figures \ref{fig8} and \ref{fig9} give informations on the Fourier
transform of the optimal pulse. One introduces the rotational
frequencies $\nu_{j+1}$ given by:
\begin{equation}
\nu_{j+1}=E_{j+1}-E_j=2B(j+1)
\end{equation}
where $E_j$ is the energy of the state $|j,0\rangle$. The target
state being associated to $j_{opt}=4$, only five rotational states
from $j=0$ to $j=4$ have to be populated by the control field. It
is thus natural to discretize the Fourier transform over the
interval $[-\nu_4,\nu_4]$. Higher frequencies do not contribute to
reach the target state. A similar behavior has been observed for
the other optimal solutions.
\begin{figure}
\includegraphics[height=1.75in]{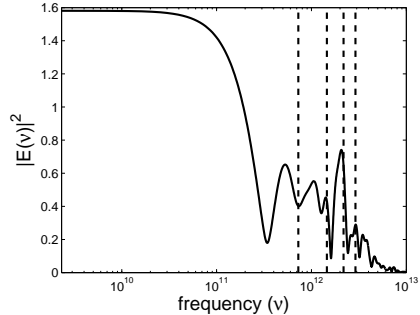}
\caption{\label{fig9} Same as Fig. \ref{fig8}a but as a function
of the frequency $\nu$. Vertical lines indicate the positions of
the frequencies $\nu_1$, $\nu_2$, $\nu_3$ and $\nu_4$.}
\end{figure}
Another example is given by Fig. \ref{fig10}. The corresponding
optimal solution obtained by the algorithm I presents rapid
unwanted oscillations. To obtain a smooth solution displayed in
Fig. \ref{fig10}b, we filter this optimal pulse in the frequency
domain. The bandwidth of the filter is chosen to cut off
frequencies higher than $\nu_4$ which produce rapid oscillations.
\begin{figure}
\includegraphics[height=1.75in]{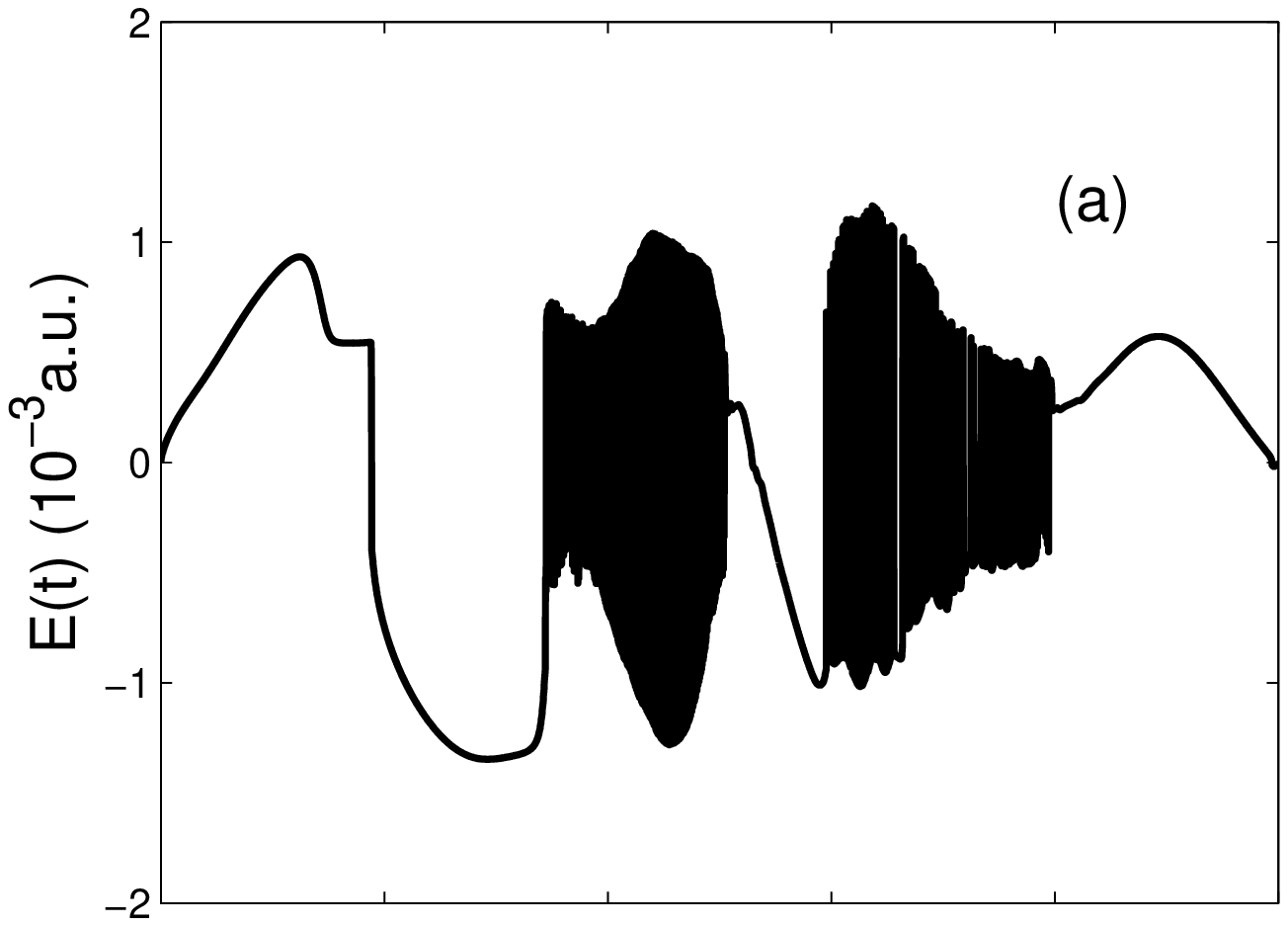}
\includegraphics[height=1.75in]{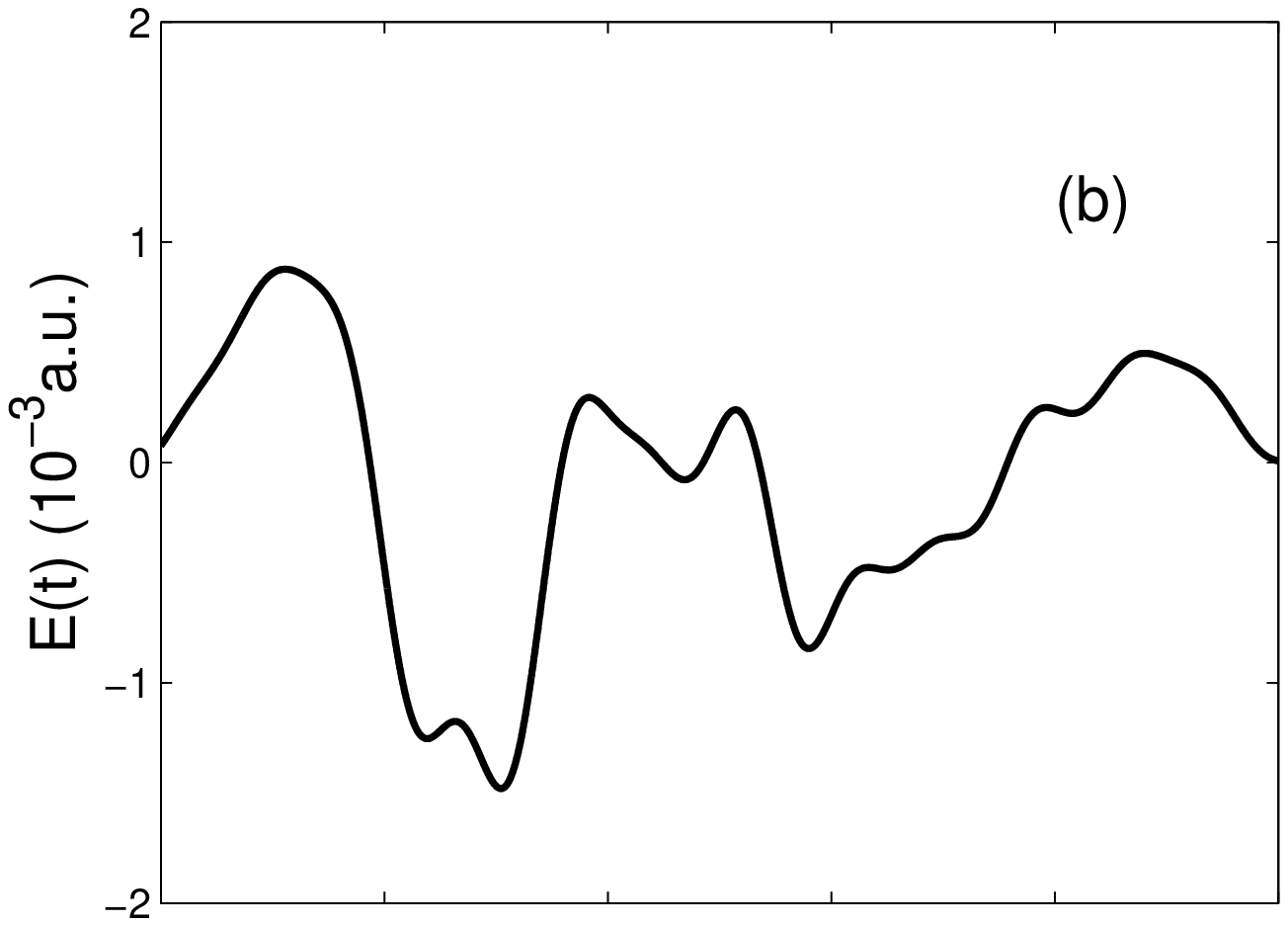}
\includegraphics[height=1.75in]{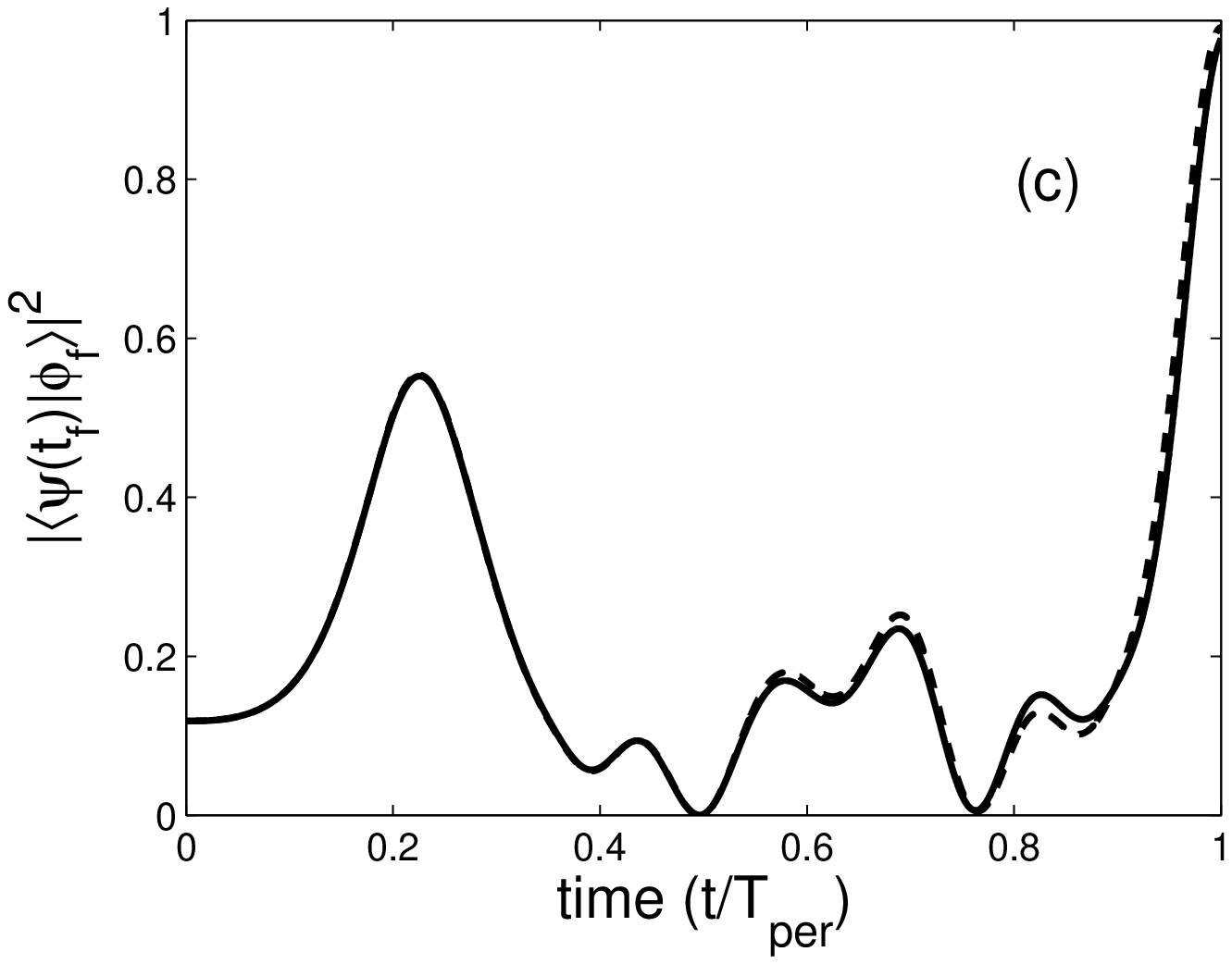}
\caption{\label{fig10} Plot as a function of time $t/T_{per}$ of
(a) the optimal (solid line) and (b) the approximate fields
(dashed line) and (c) the projection onto the target state
$|\phi_f\rangle$.}
\end{figure}
As in the first case, we also observe that the discretization does
not significantly modify the final result.
\subsection{Finite rotational temperature}\label{sec3d}
We investigate the temperature effects on the optimal solutions.
The system is described by a density matrix $\rho$ whose dynamics
is governed by the von Neumann equation. The initial density
operator $\rho(0)$ is the equilibrium density operator at
temperature $T$ which can be written
\begin{equation}
\rho(0)=\frac{1}{Z}\sum_{j=0}^{+\infty}\sum_{m=-j}^{j}e^{-Bj(j+1)/(k_BT)}|j,m\rangle\langle
j,m|
\end{equation}
where $k_B$ is the Boltzmann constant and $Z$ the partition
function. The objective of the control is to maximize the
projection of $\rho(t_f)$ onto a target state $\rho_{opt}$. We
consider here the target state introduced in Ref. \cite{sugny4}
which is unitarily equivalent to the initial mixed state
$\rho(t_i)$ and optimizes both the orientation and its duration.
We refer the reader to Ref. \cite{sugny4} for the complete
construction of $\rho_{opt}$ and for proofs of its attainability
by unitary controls. Note that the definition of $\rho_{opt}$
depends on the polarization used. We consider here the optimum for
a linear polarization. $\rho_{opt}$ can be defined as follows. The
first step consists in reducing the dimension of the Hilbert space
to a finite one $\mathcal{H}^{(j_{opt})}$ where $j_{opt}$ is the
highest $j$ for which the  corresponding rotational levels are
significantly populated. The dimension of this space depends on
the temperature and on the intensity of the field used. For $CO$
and $T=1,~ 5$ and 10 K, we have chosen $j_{opt}=4$. We denote by
$\mathcal{H}_m^{(j_{opt})}$ the subspace of
$\mathcal{H}^{(j_{opt})}$ associated to a given value of $m$. The
target state $\rho_{opt}^{(j_{opt})}$ of the control, which
therefore depends on the choice of $j_{opt}$, is given by
\begin{equation} \label{opt3}
\rho_{opt}^{(j_{opt})}=\sum_{m=-j_{opt}}^{m=j_{opt}}\sum_{k=1}^{j_{opt}-|m|+1}
\omega_k^{(m)}|\chi_k^{(m)}\rangle\langle\chi_k^{(m)}| \ ,
\end{equation}
where the $\omega_k^{(m)}$'s are the eigenvalues of $\rho(t_i)$
restricted to $\mathcal{H}_m^{(j_{opt})}$ and ordered. The vectors
$|\chi_k^{(m)}\rangle$ are the eigenvectors of the restriction of
the operator $\cos\theta$ to $\mathcal{H}_m^{(j_{opt})}$. The
vectors $|\chi_k^{(m)}\rangle$ are also ordered according to the
values of the corresponding eigenvalues.
\begin{figure}
\includegraphics[height=1.75in]{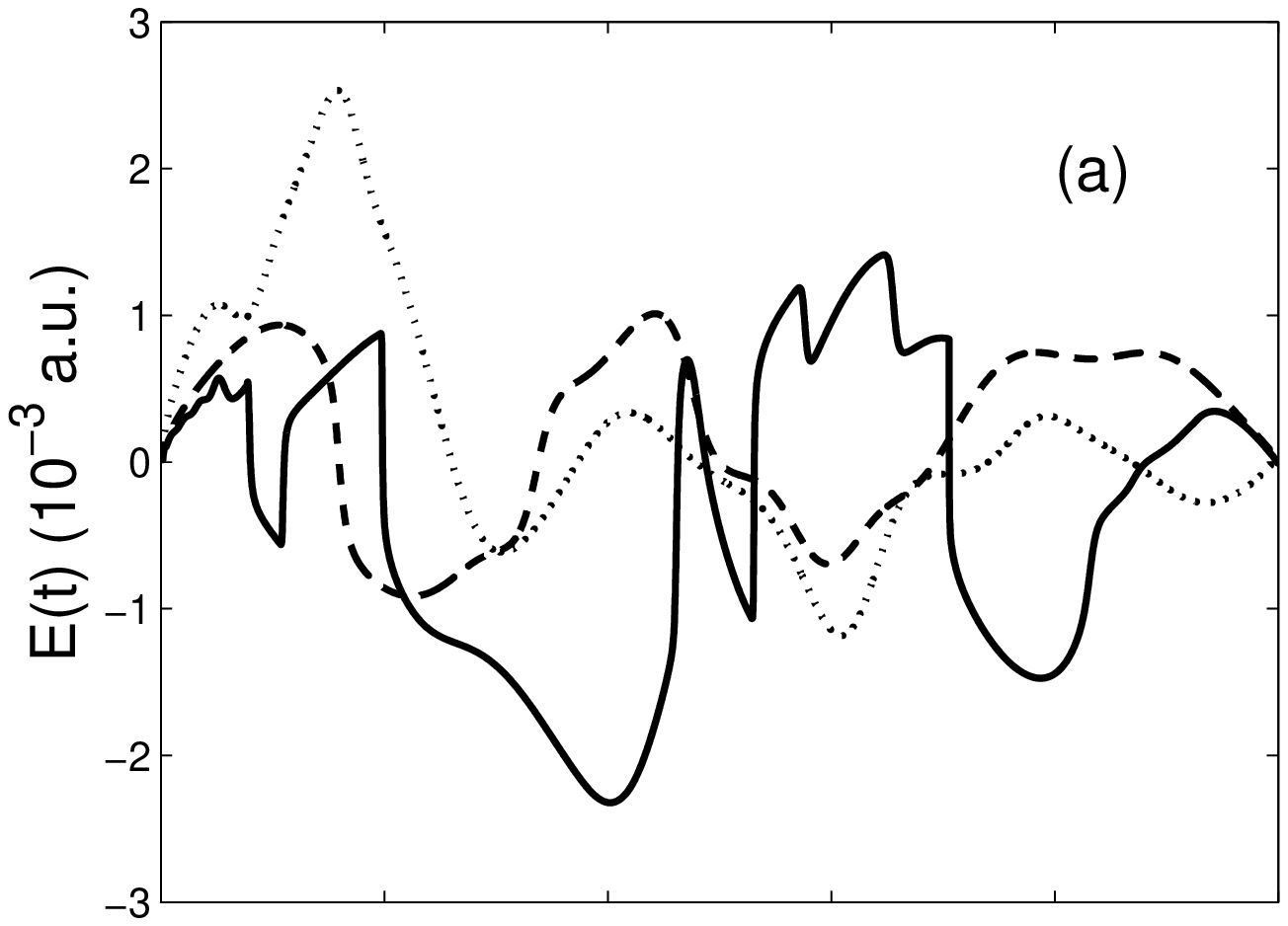}
\includegraphics[height=1.75in]{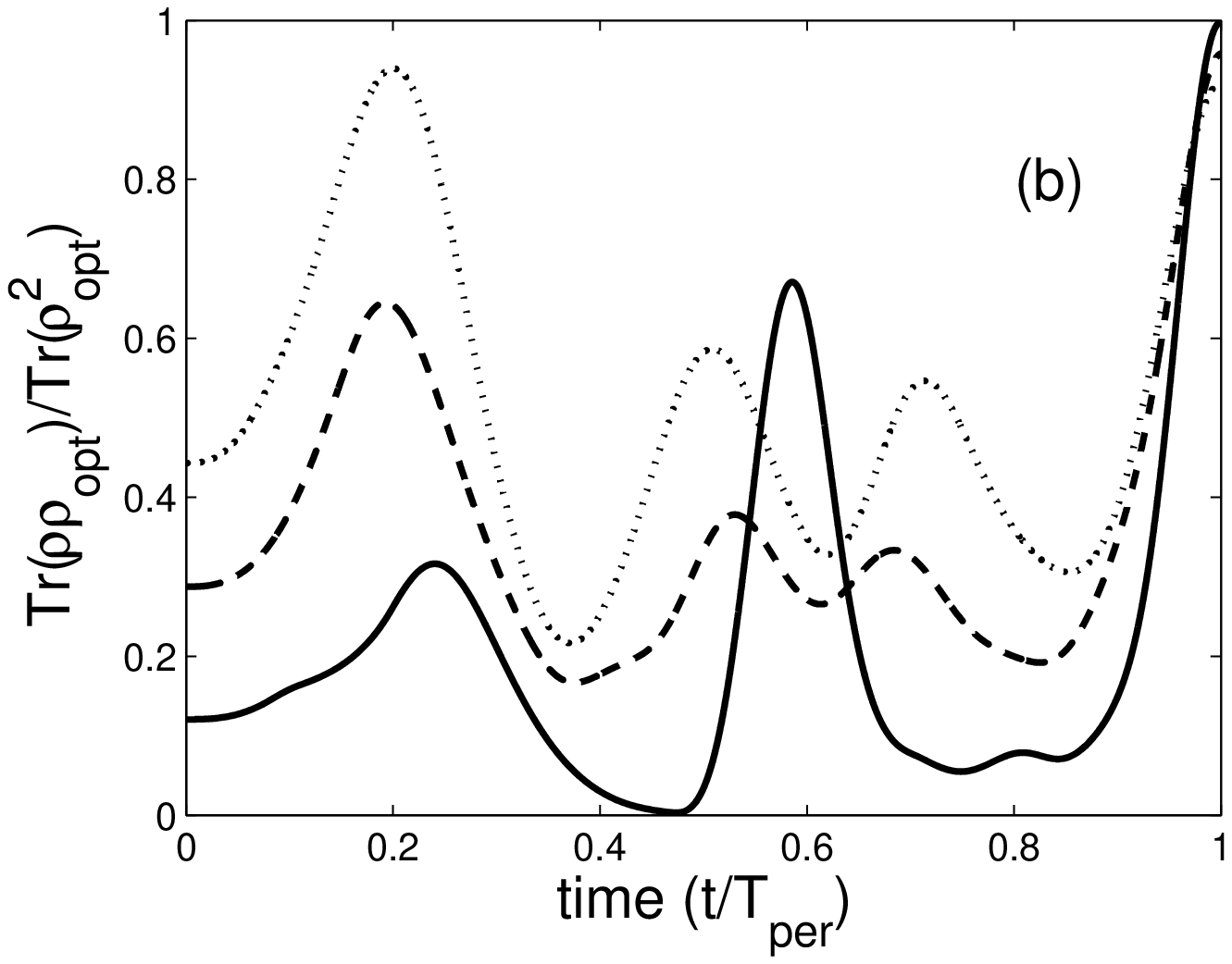}
\caption{\label{fig11} Plot as a function of the adimensional time
$t/T_{per}$ of (a) the optimal electric field and (b) the
projection onto the target state $\rho_{opt}$. Solid, dashed and
dot-dashed lines correspond respectively to $T=1~\textrm{K}$,
$T=5~\textrm{K}$ and $T=10~\textrm{K}$.}
\end{figure}

We have used the algorithm II with $n=2$ to determine the optimal
solutions. We denote by $\chi(t)$ the adjoint density matrix
state. In superoperator notations, the structure of the algorithm
is very similar to the one for pure states. The cost functional
$J$ is given by
\begin{equation}\label{opt4}
J=|\langle\langle
\rho_{opt}|\rho(t_f)\rangle\rangle|^2-\int_0^{t_f}\lambda E(t)^4dt
\end{equation}
and the augmented cost functional $\bar{J}$ reads
\begin{eqnarray}\label{opt5}
\bar{J}=& &|\langle\langle
\rho_{opt}|\rho(t_f)\rangle\rangle|^2-2\Im[\langle\langle
\rho(t_f)|\rho_{opt}\rangle\rangle\int_0^{t_f}\langle\langle\chi(t)|(i\frac{\partial}{\partial
t}-\hat{H})|\rho(t)\rangle\rangle dt]\nonumber \\
& & -\int_0^{t_f}\lambda E(t)^4dt
\end{eqnarray}
where $\langle\langle
\chi|\rho\rangle\rangle=\textrm{Tr}[\chi^\dagger\rho]$ and
$\langle\langle
\chi|M|\rho\rangle\rangle=\textrm{Tr}[\chi^\dagger[M,\rho]]$ for a
given observable $M$. $\rho(t)$ and $\chi(t)$, which satisfy the
von Neumann equation, are propagated forward and backwards with
initial condition $\rho(0)=\rho_0$ and final condition
$\chi(t_f)=\rho_{opt}$. Numerical parameters are respectively
taken to be $\lambda=6.10^4$, $\lambda=65.10^2$ and $\lambda=90$
for $T=1~\textrm{K}$, $T=5~\textrm{K}$ and $T=10~\textrm{K}$.
$\eta$ is equal to 1 in all the cases. The values of $\lambda$ are
chosen so that the total energy of the field stays approximatively
constant when the temperature is increased. The trial field is the
same for the three cases considered.

Figure \ref{fig11} presents the results obtained for three
different temperatures. We observe that the structures of the
optimal fields and of the projection onto the target state as a
function of time are very different. As expected, we note a
decrease of the final projection with increasing temperature. This
computation allows us to show the robustness with respect to
temperature of the optimal fields. For $T=10~\textrm{K}$, we still
obtain an efficient field since the final projection is of the
order of 0.9. We also point out sharp variations of the optimal
fields due to the use of a cost that is quartic in the control
field. These variations do no affect the temporal evolution of the
projection.
\section{Summary}
We have presented a new family of monotonically convergent
algorithms for the computation of the optimal control of a quantum
system interacting non-linearly with the laser field. One key for
the convergence of these algorithms is to consider costs which are
not quadratic in the field. In comparison with algorithms of Ref.
\cite{nakagami}, this allows to consider only one wave function
and one adjoint state per iteration of the algorithm whatever the
nonlinearity used. This is thus less demanding from a numerical
point of view especially when the degree of the nonlinearity is
important. As a prospect, an open question in this field is the
applicability of the present method to more complicated systems
involving, for instance, non-markovian dynamics or a
time-dependent target state \cite{time}. Special attention has to
be paid to the convergence properties and to the stability of the
method, especially when the cost is quartic in the control field.
\section*{acknowledgments}
We thank E. Hertz for many helpful discussions. We acknowledge
support from the Agence Nationale de la recherche (ANR CoMoc,
C-QUID). G.T. acknowledges support from INRIA Rocquencourt (MicMac
project) and a PICS CNRS-NSF program on quantum control.

\bibliographystyle{amsplain}

\begin{thebibliography}{48}
\expandafter\ifx\csname natexlab\endcsname\relax\def\natexlab#1{#1}\fi
\expandafter\ifx\csname bibnamefont\endcsname\relax
  \def\bibnamefont#1{#1}\fi
\expandafter\ifx\csname bibfnamefont\endcsname\relax
  \def\bibfnamefont#1{#1}\fi
\expandafter\ifx\csname citenamefont\endcsname\relax
  \def\citenamefont#1{#1}\fi
\expandafter\ifx\csname url\endcsname\relax
  \def\url#1{\texttt{#1}}\fi
\expandafter\ifx\csname urlprefix\endcsname\relax\def\urlprefix{URL }\fi
\providecommand{\bibinfo}[2]{#2}
\providecommand{\eprint}[2][]{\url{#2}}

\bibitem[{\citenamefont{1}}]{warren}
\bibinfo{author}{\bibfnamefont{W.}~\bibnamefont{Warren}},
  \bibinfo{author}{\bibfnamefont{H.}~\bibnamefont{Rabitz}}, \bibnamefont{and}
  \bibinfo{author}{\bibfnamefont{M.}~\bibnamefont{Dahleb}},
  \bibinfo{journal}{Science} \textbf{\bibinfo{volume}{259}},
  \bibinfo{pages}{1581} (\bibinfo{year}{1993}).

\bibitem[{\citenamefont{2}}]{rabitz0}
\bibinfo{author}{\bibfnamefont{H.}~\bibnamefont{Rabitz}},
  \bibinfo{author}{\bibfnamefont{R.}~\bibnamefont{de~Vivie-Riedle}},
  \bibinfo{author}{\bibfnamefont{M.}~\bibnamefont{Motzkus}}, \bibnamefont{and}
  \bibinfo{author}{\bibfnamefont{K.}~\bibnamefont{Kompa}},
  \bibinfo{journal}{Science} \textbf{\bibinfo{volume}{288}},
  \bibinfo{pages}{824} (\bibinfo{year}{2000}).

\bibitem[{\citenamefont{3}}]{nielsen}
\bibinfo{author}{\bibfnamefont{M.~A.} \bibnamefont{Nielsen}} \bibnamefont{and}
  \bibinfo{author}{\bibfnamefont{I.~L.} \bibnamefont{Chuang}},
  \emph{\bibinfo{title}{Quantum Computation and Quantum Information}}
  (\bibinfo{publisher}{Cambridge University Press},
  \bibinfo{address}{Cambridge}, \bibinfo{year}{2000}).

\bibitem[{\citenamefont{4}}]{zhu0}
\bibinfo{author}{\bibfnamefont{W.}~\bibnamefont{Zhu}} \bibnamefont{and}
  \bibinfo{author}{\bibfnamefont{H.}~\bibnamefont{Rabitz}},
  \bibinfo{journal}{J. Chem. Phys.} \textbf{\bibinfo{volume}{109}},
  \bibinfo{pages}{385} (\bibinfo{year}{1998}).

\bibitem[{\citenamefont{5}}]{maday}
\bibinfo{author}{\bibfnamefont{Y.}~\bibnamefont{Maday}} \bibnamefont{and}
  \bibinfo{author}{\bibfnamefont{G.}~\bibnamefont{Turinici}},
  \bibinfo{journal}{J. Chem. Phys.} \textbf{\bibinfo{volume}{118}},
  \bibinfo{pages}{8191} (\bibinfo{year}{2003}).

\bibitem[{\citenamefont{6}}]{schirmer}
\bibinfo{author}{\bibfnamefont{S.}~\bibnamefont{Schirmer}},
  \bibinfo{author}{\bibfnamefont{M.}~\bibnamefont{Girardeau}},
  \bibnamefont{and} \bibinfo{author}{\bibfnamefont{J.}~\bibnamefont{Leahy}},
  \bibinfo{journal}{Phys. Rev. A} \textbf{\bibinfo{volume}{61}},
  \bibinfo{pages}{012101} (\bibinfo{year}{2000}).

\bibitem[{\citenamefont{7}}]{ohtsuki1}
\bibinfo{author}{\bibfnamefont{Y.}~\bibnamefont{Ohtsuki}},
  \bibinfo{author}{\bibfnamefont{W.}~\bibnamefont{Zhu}}, \bibnamefont{and}
  \bibinfo{author}{\bibfnamefont{H.}~\bibnamefont{Rabitz}},
  \bibinfo{journal}{J. Chem. Phys.} \textbf{\bibinfo{volume}{110}},
  \bibinfo{pages}{9825} (\bibinfo{year}{1999}).

\bibitem[{\citenamefont{8}}]{ohtsuki2}
\bibinfo{author}{\bibfnamefont{Y.}~\bibnamefont{Ohtsuki}},
  \bibinfo{author}{\bibfnamefont{G.}~\bibnamefont{Turinici}}, \bibnamefont{and}
  \bibinfo{author}{\bibfnamefont{H.}~\bibnamefont{Rabitz}},
  \bibinfo{journal}{J. Chem. Phys.} \textbf{\bibinfo{volume}{120}},
  \bibinfo{pages}{5509} (\bibinfo{year}{2004}).

\bibitem[{\citenamefont{9}}]{sugnybook}
\bibinfo{author}{\bibfnamefont{B.}~\bibnamefont{Bonnard}} \bibnamefont{and}
  \bibinfo{author}{\bibfnamefont{D.}~\bibnamefont{Sugny}},
  \emph{\bibinfo{title}{Optimal control with applications in space and quantum
  dynamics}} (\bibinfo{publisher}{AIMS Applied Maths},
  \bibinfo{address}{submitted}, \bibinfo{year}{2008}).

\bibitem[{\citenamefont{10}}]{sugnynew}
\bibinfo{author}{\bibfnamefont{D.}~\bibnamefont{Sugny}},
  \bibinfo{author}{\bibfnamefont{C.}~\bibnamefont{Kontz}}, \bibnamefont{and}
  \bibinfo{author}{\bibfnamefont{H.}~\bibnamefont{Jauslin}},
  \bibinfo{journal}{Phys. Rev. A} \textbf{\bibinfo{volume}{76}},
  \bibinfo{pages}{023419} (\bibinfo{year}{2007}{\natexlab{a}}).

\bibitem[{\citenamefont{11}}]{gross}
\bibinfo{author}{\bibfnamefont{J.}~\bibnamefont{Werschnik}} \bibnamefont{and}
  \bibinfo{author}{\bibfnamefont{E.~K.~U.} \bibnamefont{Gross}},
  \bibinfo{journal}{J. Phys. B} \textbf{\bibinfo{volume}{40}},
  \bibinfo{pages}{R175} (\bibinfo{year}{2007}).

\bibitem[{\citenamefont{12}}]{nakagami}
\bibinfo{author}{\bibfnamefont{Y.}~\bibnamefont{Ohtsuki}} \bibnamefont{and}
  \bibinfo{author}{\bibfnamefont{K.}~\bibnamefont{Nakagami}},
  \bibinfo{journal}{Phys. Rev. A} \textbf{\bibinfo{volume}{77}},
  \bibinfo{pages}{033414} (\bibinfo{year}{2008}).

\bibitem[{\citenamefont{13}}]{potz1}
\bibinfo{author}{\bibfnamefont{H.}~\bibnamefont{Jirari}} \bibnamefont{and}
  \bibinfo{author}{\bibfnamefont{W.}~\bibnamefont{Potz}},
  \bibinfo{journal}{Phys. Rev. A} \textbf{\bibinfo{volume}{72}},
  \bibinfo{pages}{013409} (\bibinfo{year}{2005}).

\bibitem[{\citenamefont{14}}]{potz2}
\bibinfo{author}{\bibfnamefont{M.}~\bibnamefont{Wenin}} \bibnamefont{and}
  \bibinfo{author}{\bibfnamefont{W.}~\bibnamefont{Potz}},
  \bibinfo{journal}{Phys. Rev. A} \textbf{\bibinfo{volume}{74}},
  \bibinfo{pages}{022319} (\bibinfo{year}{2006}).

\bibitem[{\citenamefont{15}}]{bryson}
\bibinfo{author}{\bibfnamefont{A.~E.} \bibnamefont{Bryson}} \bibnamefont{and}
  \bibinfo{author}{\bibfnamefont{Y.}~\bibnamefont{Ho}},
  \emph{\bibinfo{title}{Applied optimal control}}
  (\bibinfo{publisher}{Hemisphere Publishing Corporation},
  \bibinfo{address}{Washington}, \bibinfo{year}{1975}).

\bibitem[{\citenamefont{16}}]{tannor}
\bibinfo{author}{\bibfnamefont{R.}~\bibnamefont{Kosloff}},
  \bibinfo{author}{\bibfnamefont{S.}~\bibnamefont{Rice}},
  \bibinfo{author}{\bibfnamefont{P.}~\bibnamefont{Gaspard}},
  \bibinfo{author}{\bibfnamefont{S.}~\bibnamefont{Tersigni}}, \bibnamefont{and}
  \bibinfo{author}{\bibfnamefont{D.}~\bibnamefont{Tannor}},
  \bibinfo{journal}{Chem. Phys.} \textbf{\bibinfo{volume}{139}},
  \bibinfo{pages}{201} (\bibinfo{year}{1989}).

\bibitem[{\citenamefont{17}}]{zhu1}
\bibinfo{author}{\bibfnamefont{W.}~\bibnamefont{Zhu}},
  \bibinfo{author}{\bibfnamefont{J.}~\bibnamefont{Botina}}, \bibnamefont{and}
  \bibinfo{author}{\bibfnamefont{H.}~\bibnamefont{Rabitz}},
  \bibinfo{journal}{J. Chem. Phys.} \textbf{\bibinfo{volume}{108}},
  \bibinfo{pages}{1953} (\bibinfo{year}{1998}).

\bibitem[{\citenamefont{18}}]{zhu}
\bibinfo{author}{\bibfnamefont{W.}~\bibnamefont{Zhu}} \bibnamefont{and}
  \bibinfo{author}{\bibfnamefont{H.}~\bibnamefont{Rabitz}},
  \bibinfo{journal}{J. Chem. Phys.} \textbf{\bibinfo{volume}{110}},
  \bibinfo{pages}{7142} (\bibinfo{year}{1999}).

\bibitem[{\citenamefont{19}}]{sugny7}
\bibinfo{author}{\bibfnamefont{M.}~\bibnamefont{Ndong}},
  \bibinfo{author}{\bibfnamefont{L.}~\bibnamefont{Bomble}},
  \bibinfo{author}{\bibfnamefont{D.}~\bibnamefont{Sugny}},
  \bibinfo{author}{\bibfnamefont{Y.}~\bibnamefont{Justum}}, \bibnamefont{and}
  \bibinfo{author}{\bibfnamefont{M.}~\bibnamefont{Desouter-Lecomte}},
  \bibinfo{journal}{Phys. Rev. A} \textbf{\bibinfo{volume}{76}},
  \bibinfo{pages}{043424} (\bibinfo{year}{2007}).

\bibitem[{\citenamefont{20}}]{bifurcating}
\bibinfo{author}{\bibfnamefont{D.}~\bibnamefont{Sugny}},
  \bibinfo{author}{\bibfnamefont{C.}~\bibnamefont{Kontz}},
  \bibinfo{author}{\bibfnamefont{M.}~\bibnamefont{Ndong}},
  \bibinfo{author}{\bibfnamefont{Y.}~\bibnamefont{Justum}},
  \bibinfo{author}{\bibfnamefont{G.}~\bibnamefont{Dives}}, \bibnamefont{and}
  \bibinfo{author}{\bibfnamefont{M.}~\bibnamefont{Desouter-Lecomte}},
  \bibinfo{journal}{Phys. Rev. A} \textbf{\bibinfo{volume}{74}},
  \bibinfo{pages}{043419} (\bibinfo{year}{2006}{\natexlab{a}}).

\bibitem[{\citenamefont{21}}]{sugny2}
\bibinfo{author}{\bibfnamefont{D.}~\bibnamefont{Sugny}},
  \bibinfo{author}{\bibfnamefont{M.}~\bibnamefont{Ndong}},
  \bibinfo{author}{\bibfnamefont{D.}~\bibnamefont{Lauvergnat}},
  \bibinfo{author}{\bibfnamefont{Y.}~\bibnamefont{Justum}}, \bibnamefont{and}
  \bibinfo{author}{\bibfnamefont{M.}~\bibnamefont{Desouter-Lecomte}},
  \bibinfo{journal}{J. Photochem. Photobiol. A : Chemistry}
  \textbf{\bibinfo{volume}{190}}, \bibinfo{pages}{359}
  (\bibinfo{year}{2007}{\natexlab{b}}).

\bibitem[{\citenamefont{22}}]{salomon}
\bibinfo{author}{\bibfnamefont{J.}~\bibnamefont{Salomon}},
  \bibinfo{author}{\bibfnamefont{C.~M.} \bibnamefont{Dion}}, \bibnamefont{and}
  \bibinfo{author}{\bibfnamefont{G.}~\bibnamefont{Turinici}},
  \bibinfo{journal}{J. Chem. Phys.} \textbf{\bibinfo{volume}{123}},
  \bibinfo{pages}{144310} (\bibinfo{year}{2005}).

\bibitem[{\citenamefont{23}}]{saalfrank}
\bibinfo{author}{\bibfnamefont{Y.}~\bibnamefont{Ohtsuki}},
  \bibinfo{author}{\bibfnamefont{Y.}~\bibnamefont{Teranishi}},
  \bibinfo{author}{\bibfnamefont{P.}~\bibnamefont{Saalfrank}},
  \bibinfo{author}{\bibfnamefont{G.}~\bibnamefont{Turinici}}, \bibnamefont{and}
  \bibinfo{author}{\bibfnamefont{H.}~\bibnamefont{Rabitz}},
  \bibinfo{journal}{Phys. Rev. A} \textbf{\bibinfo{volume}{75}},
  \bibinfo{pages}{033407} (\bibinfo{year}{2007}).

\bibitem[{\citenamefont{24}}]{friedrich}
\bibinfo{author}{\bibfnamefont{B.}~\bibnamefont{Friedrich}} \bibnamefont{and}
  \bibinfo{author}{\bibfnamefont{D.}~\bibnamefont{Herschbach}},
  \bibinfo{journal}{Phys. Rev. Lett.} \textbf{\bibinfo{volume}{74}},
  \bibinfo{pages}{4623} (\bibinfo{year}{1995}).

\bibitem[{\citenamefont{25}}]{stolow1}
\bibinfo{author}{\bibfnamefont{B.~J.} \bibnamefont{Sussman}},
  \bibinfo{author}{\bibfnamefont{M.~Y.} \bibnamefont{Ivanov}},
  \bibnamefont{and} \bibinfo{author}{\bibfnamefont{A.}~\bibnamefont{Stolow}},
  \bibinfo{journal}{Phys. Rev. A} \textbf{\bibinfo{volume}{71}},
  \bibinfo{pages}{R051401} (\bibinfo{year}{2005}).

\bibitem[{\citenamefont{26}}]{stolow2}
\bibinfo{author}{\bibfnamefont{J.~G.} \bibnamefont{Underwood}},
  \bibinfo{author}{\bibfnamefont{M.}~\bibnamefont{Spanner}},
  \bibinfo{author}{\bibfnamefont{M.~Y.} \bibnamefont{Ivanov}},
  \bibinfo{author}{\bibfnamefont{J.}~\bibnamefont{Mottershead}},
  \bibinfo{author}{\bibfnamefont{B.~J.} \bibnamefont{Sussman}},
  \bibnamefont{and} \bibinfo{author}{\bibfnamefont{A.}~\bibnamefont{Stolow}},
  \bibinfo{journal}{Phys. Rev. Lett.} \textbf{\bibinfo{volume}{90}},
  \bibinfo{pages}{223001} (\bibinfo{year}{2003}).

\bibitem[{\citenamefont{27}}]{tehini}
\bibinfo{author}{\bibfnamefont{R.}~\bibnamefont{Tehini}} \bibnamefont{and}
  \bibinfo{author}{\bibfnamefont{D.}~\bibnamefont{Sugny}},
  \bibinfo{journal}{Phys. Rev. A} \textbf{\bibinfo{volume}{77}},
  \bibinfo{pages}{023407} (\bibinfo{year}{2008}).

\bibitem[{\citenamefont{28}}]{seideman}
\bibinfo{author}{\bibfnamefont{H.}~\bibnamefont{Stapelfeldt}} \bibnamefont{and}
  \bibinfo{author}{\bibfnamefont{T.}~\bibnamefont{Seideman}},
  \bibinfo{journal}{Rev. Mod. Phys.} \textbf{\bibinfo{volume}{75}},
  \bibinfo{pages}{543} (\bibinfo{year}{2003}).

\bibitem[{\citenamefont{29}}]{seideman1}
\bibinfo{author}{\bibfnamefont{T.}~\bibnamefont{Seideman}} \bibnamefont{and}
  \bibinfo{author}{\bibfnamefont{E.}~\bibnamefont{Hamilton}},
  \bibinfo{journal}{Adv. At. Mol. Opt. Phys.} \textbf{\bibinfo{volume}{52}},
  \bibinfo{pages}{289} (\bibinfo{year}{2006}).

\bibitem[{\citenamefont{30}}]{oberwolfach}
\bibinfo{author}{\bibfnamefont{G.}~\bibnamefont{Turinici}}, in
  \emph{\bibinfo{booktitle}{Control of coupled partial differential equations}}
  (\bibinfo{publisher}{Birkh\"auser}, \bibinfo{address}{Basel},
  \bibinfo{year}{2007}), vol. \bibinfo{volume}{155} of
  \emph{\bibinfo{series}{Internat. Ser. Numer. Math.}}, pp.
  \bibinfo{pages}{293--309}.

\bibitem[{\citenamefont{31}}]{sugny6}
\bibinfo{author}{\bibfnamefont{D.}~\bibnamefont{Sugny}},
  \bibinfo{author}{\bibfnamefont{A.}~\bibnamefont{Keller}},
  \bibinfo{author}{\bibfnamefont{O.}~\bibnamefont{Atabek}},
  \bibinfo{author}{\bibfnamefont{D.}~\bibnamefont{Daems}},
  \bibinfo{author}{\bibfnamefont{C.~M.} \bibnamefont{Dion}},
  \bibinfo{author}{\bibfnamefont{S.}~\bibnamefont{Gu\'erin}}, \bibnamefont{and}
  \bibinfo{author}{\bibfnamefont{H.~R.} \bibnamefont{Jauslin}},
  \bibinfo{journal}{Phys. Rev. A} \textbf{\bibinfo{volume}{69}},
  \bibinfo{pages}{033402} (\bibinfo{year}{2004}{\natexlab{a}}).

\bibitem[{\citenamefont{32}}]{sugny3}
\bibinfo{author}{\bibfnamefont{D.}~\bibnamefont{Sugny}},
  \bibinfo{author}{\bibfnamefont{A.}~\bibnamefont{Keller}},
  \bibinfo{author}{\bibfnamefont{O.}~\bibnamefont{Atabek}},
  \bibinfo{author}{\bibfnamefont{D.}~\bibnamefont{Daems}},
  \bibinfo{author}{\bibfnamefont{C.~M.} \bibnamefont{Dion}},
  \bibinfo{author}{\bibfnamefont{S.}~\bibnamefont{Gu\'erin}}, \bibnamefont{and}
  \bibinfo{author}{\bibfnamefont{H.~R.} \bibnamefont{Jauslin}},
  \bibinfo{journal}{Phys. Rev. A} \textbf{\bibinfo{volume}{71}},
  \bibinfo{pages}{063402} (\bibinfo{year}{2005}{\natexlab{a}}).

\bibitem[{\citenamefont{33}}]{sugny4}
\bibinfo{author}{\bibfnamefont{D.}~\bibnamefont{Sugny}},
  \bibinfo{author}{\bibfnamefont{A.}~\bibnamefont{Keller}},
  \bibinfo{author}{\bibfnamefont{O.}~\bibnamefont{Atabek}},
  \bibinfo{author}{\bibfnamefont{D.}~\bibnamefont{Daems}},
  \bibinfo{author}{\bibfnamefont{C.~M.} \bibnamefont{Dion}},
  \bibinfo{author}{\bibfnamefont{S.}~\bibnamefont{Gu\'erin}}, \bibnamefont{and}
  \bibinfo{author}{\bibfnamefont{H.~R.} \bibnamefont{Jauslin}},
  \bibinfo{journal}{Phys. Rev. A} \textbf{\bibinfo{volume}{72}},
  \bibinfo{pages}{032704} (\bibinfo{year}{2005}{\natexlab{b}}).

\bibitem[{\citenamefont{34}}]{sugny1}
\bibinfo{author}{\bibfnamefont{D.}~\bibnamefont{Sugny}},
  \bibinfo{author}{\bibfnamefont{C.}~\bibnamefont{Kontz}}, \bibnamefont{and}
  \bibinfo{author}{\bibfnamefont{H.}~\bibnamefont{Jauslin}},
  \bibinfo{journal}{Phys. Rev. A} \textbf{\bibinfo{volume}{74}},
  \bibinfo{pages}{053411} (\bibinfo{year}{2006}{\natexlab{b}}).

\bibitem[{\citenamefont{35}}]{judson}
\bibinfo{author}{\bibfnamefont{R.~S.} \bibnamefont{Judson}} \bibnamefont{and}
  \bibinfo{author}{\bibfnamefont{H.}~\bibnamefont{Rabitz}},
  \bibinfo{journal}{Phys. Rev. Lett.} \textbf{\bibinfo{volume}{68}},
  \bibinfo{pages}{1500} (\bibinfo{year}{1992}).

\bibitem[{\citenamefont{36}}]{science1}
\bibinfo{author}{\bibfnamefont{A.}~\bibnamefont{Assion}},
  \bibinfo{author}{\bibfnamefont{T.}~\bibnamefont{Baumer}},
  \bibinfo{author}{\bibfnamefont{M.}~\bibnamefont{Bergt}},
  \bibinfo{author}{\bibfnamefont{T.}~\bibnamefont{Brixner}},
  \bibinfo{author}{\bibfnamefont{B.}~\bibnamefont{Kiefer}},
  \bibinfo{author}{\bibfnamefont{V.}~\bibnamefont{Seyfried}},
  \bibinfo{author}{\bibfnamefont{M.}~\bibnamefont{Strehle}}, \bibnamefont{and}
  \bibinfo{author}{\bibfnamefont{G.}~\bibnamefont{Gerber}},
  \bibinfo{journal}{Science} \textbf{\bibinfo{volume}{282}},
  \bibinfo{pages}{919} (\bibinfo{year}{1998}).

\bibitem[{\citenamefont{37}}]{science2}
\bibinfo{author}{\bibfnamefont{R.~J.} \bibnamefont{Levis}},
  \bibinfo{author}{\bibfnamefont{G.~M.} \bibnamefont{Menkir}},
  \bibnamefont{and} \bibinfo{author}{\bibfnamefont{H.}~\bibnamefont{Rabitz}},
  \bibinfo{journal}{Science} \textbf{\bibinfo{volume}{292}},
  \bibinfo{pages}{709} (\bibinfo{year}{2001}).

\bibitem[{\citenamefont{38}}]{science3}
\bibinfo{author}{\bibfnamefont{C.}~\bibnamefont{Daniel}},
  \bibinfo{author}{\bibfnamefont{J.}~\bibnamefont{Full}},
  \bibinfo{author}{\bibfnamefont{L.}~\bibnamefont{Gonz\`alez}},
  \bibinfo{author}{\bibfnamefont{C.}~\bibnamefont{Lupulescu}},
  \bibinfo{author}{\bibfnamefont{J.}~\bibnamefont{Manz}},
  \bibinfo{author}{\bibfnamefont{A.}~\bibnamefont{Merli}},
  \bibinfo{author}{\bibfnamefont{S.}~\bibnamefont{Vadja}}, \bibnamefont{and}
  \bibinfo{author}{\bibfnamefont{L.}~\bibnamefont{Woste}},
  \bibinfo{journal}{Science} \textbf{\bibinfo{volume}{299}},
  \bibinfo{pages}{536} (\bibinfo{year}{2003}).

\bibitem[{\citenamefont{39}}]{shir}
\bibinfo{author}{\bibfnamefont{O.~M.} \bibnamefont{Shir}},
  \bibinfo{author}{\bibfnamefont{V.}~\bibnamefont{Beltrani}},
  \bibinfo{author}{\bibfnamefont{T.}~\bibnamefont{Back}},
  \bibinfo{author}{\bibfnamefont{H.}~\bibnamefont{Rabitz}}, \bibnamefont{and}
  \bibinfo{author}{\bibfnamefont{M.}~\bibnamefont{Vrakking}},
  \bibinfo{journal}{J. Phys. B} \textbf{\bibinfo{volume}{41}},
  \bibinfo{pages}{074021} (\bibinfo{year}{2008}).

\bibitem[{\citenamefont{40}}]{hertz1}
\bibinfo{author}{\bibfnamefont{E.}~\bibnamefont{Hertz}},
  \bibinfo{author}{\bibfnamefont{A.}~\bibnamefont{Rouz\'ee}},
  \bibinfo{author}{\bibfnamefont{S.}~\bibnamefont{Gu\'erin}},
  \bibinfo{author}{\bibfnamefont{B.}~\bibnamefont{Lavorel}}, \bibnamefont{and}
  \bibinfo{author}{\bibfnamefont{O.}~\bibnamefont{Faucher}},
  \bibinfo{journal}{Phys. Rev. A} \textbf{\bibinfo{volume}{75}},
  \bibinfo{pages}{031403} (\bibinfo{year}{2007}).

\bibitem[{\citenamefont{41}}]{hertz2}
\bibinfo{author}{\bibfnamefont{A.}~\bibnamefont{Rouz\'ee}},
  \bibinfo{author}{\bibfnamefont{E.}~\bibnamefont{Hertz}},
  \bibinfo{author}{\bibfnamefont{B.}~\bibnamefont{Lavorel}}, \bibnamefont{and}
  \bibinfo{author}{\bibfnamefont{O.}~\bibnamefont{Faucher}},
  \bibinfo{journal}{J. Phys. B} \textbf{\bibinfo{volume}{41}},
  \bibinfo{pages}{074002} (\bibinfo{year}{2008}).

\bibitem[{\citenamefont{42}}]{sundermann}
\bibinfo{author}{\bibfnamefont{K.}~\bibnamefont{Sundermann}} \bibnamefont{and}
  \bibinfo{author}{\bibfnamefont{R.}~\bibnamefont{de~Vivie-Riedle}},
  \bibinfo{journal}{J. Chem. Phys.} \textbf{\bibinfo{volume}{110}},
  \bibinfo{pages}{1896} (\bibinfo{year}{1999}).

\bibitem[{\citenamefont{43}}]{seiddiss3}
\bibinfo{author}{\bibfnamefont{A.}~\bibnamefont{Pelzer}},
  \bibinfo{author}{\bibfnamefont{S.}~\bibnamefont{Ramakrishna}},
  \bibnamefont{and} \bibinfo{author}{\bibfnamefont{T.}~\bibnamefont{Seideman}},
  \bibinfo{journal}{J. Chem. Phys.} \textbf{\bibinfo{volume}{126}},
  \bibinfo{pages}{034503} (\bibinfo{year}{2007}).

\bibitem[{\citenamefont{44}}]{sugnykick}
\bibinfo{author}{\bibfnamefont{D.}~\bibnamefont{Sugny}},
  \bibinfo{author}{\bibfnamefont{A.}~\bibnamefont{Keller}},
  \bibinfo{author}{\bibfnamefont{O.}~\bibnamefont{Atabek}},
  \bibinfo{author}{\bibfnamefont{D.}~\bibnamefont{Daems}},
  \bibinfo{author}{\bibfnamefont{S.}~\bibnamefont{Gu\'erin}}, \bibnamefont{and}
  \bibinfo{author}{\bibfnamefont{H.~R.} \bibnamefont{Jauslin}},
  \bibinfo{journal}{Phys. Rev. A} \textbf{\bibinfo{volume}{69}},
  \bibinfo{pages}{043407} (\bibinfo{year}{2004}{\natexlab{b}}).

\bibitem[{\citenamefont{45}}]{kanai}
\bibinfo{author}{\bibfnamefont{T.}~\bibnamefont{Kanai}} \bibnamefont{and}
  \bibinfo{author}{\bibfnamefont{H.}~\bibnamefont{Sakai}}, \bibinfo{journal}{J.
  Chem. Phys.} \textbf{\bibinfo{volume}{115}}, \bibinfo{pages}{5492}
  (\bibinfo{year}{2001}).

\bibitem[{\citenamefont{46}}]{sekino}
\bibinfo{author}{\bibfnamefont{H.}~\bibnamefont{Sekino}} \bibnamefont{and}
  \bibinfo{author}{\bibfnamefont{R.~J.} \bibnamefont{Bartlett}},
  \bibinfo{journal}{J. Chem. Phys.} \textbf{\bibinfo{volume}{98}},
  \bibinfo{pages}{3022} (\bibinfo{year}{1993}).

\bibitem[{\citenamefont{47}}]{maroulis}
\bibinfo{author}{\bibfnamefont{G.}~\bibnamefont{Maroulis}},
  \bibinfo{journal}{J. Phys. Chem.} \textbf{\bibinfo{volume}{100}},
  \bibinfo{pages}{13466} (\bibinfo{year}{1996}).

\bibitem[{\citenamefont{48}}]{time}
\bibinfo{author}{\bibfnamefont{I.}~\bibnamefont{Serban}},
  \bibinfo{author}{\bibfnamefont{J.}~\bibnamefont{Werschnik}},
  \bibnamefont{and} \bibinfo{author}{\bibfnamefont{E.~K.~U.}
  \bibnamefont{Gross}}, \bibinfo{journal}{Phys. Rev. A}
  \textbf{\bibinfo{volume}{71}}, \bibinfo{pages}{053810}
  (\bibinfo{year}{2005}).

\end{thebibliography}

\end{document}